\documentclass[structabstract]{aa}
\usepackage{txfonts}
\usepackage{epsf}
\usepackage{graphicx}
\usepackage{natbib}
\bibpunct{(}{)}{;}{a}{}{,} 

\def\lya{Ly$\alpha$~}
\def\ha{H$\alpha$~}
\def\hb{H$\beta$~}
\def\xha{H$\alpha$}
\def\xhb{H$\beta$}

\def\hahb{\xha/\hb}
\def\xhahb{\xha/\xhb}
\def\wha{EW(H$\alpha$)~}
\def\whb{EW(H$\beta$)~}
\def\xwha{EW(H$\alpha$)}

\def\o5{[\ion{O}{iii}]$\lambda$5007}
\def\03{[\ion{O}{ii}]$\lambda$3727}
\def\xo5{[\ion{O}{iii}]$\lambda$5007~}
\def\xo3{[\ion{O}{ii}]$\lambda$3727~}

\def\sma{$\cal M_\odot$}

\def\ltapprox{\mathrel{\raise.3ex\hbox{$<$}\mkern-14mu
             \lower0.6ex\hbox{$\sim$}}}

\begin{document}
\title{Lyman continuum leaking galaxies - \\Search strategies and local candidates\thanks{Based on observations made with ESO Telescopes at the La Silla or Paranal Observatories under programme ID 080.B-0899(A) }}

\author{Nils Bergvall\inst{1}
\and Elisabet Leitet\inst{1}
\and Erik Zackrisson\inst{2} 
\and Thomas Marquart\inst{1}}

\institute{{Department of Physics and Astronomy, Uppsala University, Box 515, SE-751 20 Uppsala,
SWEDEN}\\
\email{nils.bergvall@physics.uu.se, elisabet.leitet@physics.uu.se, thomas.marquart@physics.uu.se}
\and{Department of Astronomy, Stockholm University, Oscar Klein Center, AlbaNova, Stockholm SE-106 91, Sweden}\\
\email{ez@astro.su.se}}

\date{Submitted }

\abstract{Star-forming dwarf galaxies leaking Lyman-continuum (LyC) radiation may have played an important role in the reionization of the Universe. Local galaxies exhibiting LyC leakage could shed light on the escape mechanisms, but so far only two such cases have been identified. Here, we argue that this meager harvest may be caused by unsuitable selection criteria, one of which is a high equivalent width in H$\alpha$ emission. Such a criterion will guarantee a high production of LyC photons but also indicates a high column density in the neutral gas, effectively ruling out LyC escape.}
{We want to investigate whether the lack of local LyC emitters can be caused in part by biased selection criteria, and we present a
novel method of selecting targets with high escape fractions. By applying
these criteria, we assemble a sample of observation targets to study
their basic properties.}
{We introduce a new selection strategy where the potential LyC leakers are selected by their blue colours and $weak$ emission lines. We also take a closer look at the properties of  8 LyC leaking candidates at $z \sim$ 0.03, selected from the Sloan Digital Sky Survey (SDSS) which we observe with ESO/NTT in broadband $B$ and  \xha. }
{We find that 7 of the 8 target galaxies are involved in interaction with neighbours or show signs of mergers. In 7 cases the young stellar population is clearly displaced relative to the main body of these galaxies. In about half of our targets the absorption spectra show young post-starburst signatures.  
The scale lengths in \ha are typically 30\% smaller than those of the stellar continua, which is characteristic of galaxies influenced by ram pressure stripping.  We tentatively identify a few conditions favourable for leakage: 1)  the combined effects of ram pressure stripping with supernova winds from young stars formed in the front, 2) merger events that increase the star formation rate and displace stars from gas, 3) starbursts in the centres of post-starburst galaxies, and 4) a low dust content. 
}
{}

\keywords{Galaxies: intergalactic medium - starburst - fundamental parameters - evolution, Cosmology: diffuse radiation, Ultraviolet: galaxies }

\authorrunning{Bergvall et al.}
\titlerunning{Lyman continuum leaking galaxies}
\maketitle

\section{Introduction}
After cosmic recombination the "dark ages" followed. It is an epoch during which no astronomical light sources existed and the gas in the universe remained neutral. A few hundred million years later, the universe underwent reionization. The seven-year WMAP data suggests that this process started at $z\gtrsim 10$ \citep{2011ApJS..192...18K}, and observations of Gunn-Peterson troughs in the spectra of high-redshift quasars places the completion of this process at $z\approx 6$ \citep[e.g.][]{2001AJ....121...54F, 2006AJ....132..117F}. The sources responsible for reionization still, however, need to be identified. 

Accreting black holes and population III stars could have contributed to the LyC budget during the reionization epoch \citep[e.g.][]{2004ApJ...604..484M,2009ApJ...694..879T,2009ApJ...703.2113V,2011A&A...528A.149M}, but galaxies of low to intermediate masses seem to be the top candidates. Star formation models and high-z observations \citep[e.g][]{2007ApJ...663...10S,2011Natur.469..504B} imply that galaxies were already in place at $z\approx 10$ and that starburst dwarf galaxies may have played a dominant role. Cosmological simulations  give further support to the importance of galaxies during the reionization epoch, even though predictions vary as to the exact mass scales of the objects giving the largest LyC contribution \citep[e.g][]{2008ApJ...672..765G,2009ApJ...693..984W,2010ApJ...710.1239R,2011MNRAS.412..411Y}.

Most attempts to detect the escape of hydrogen-ionizing (i.e. Lyman continuum, hereafter LyC) radiation from local galaxies have failed \citep[e.g.][]{1995ApJ...454L..19L,2001A&A...375..805D,2009ApJS..181..272G}. Even though a few galaxies, including our own Milky Way \citep{1999ApJ...510L..33B}, display indirect signatures of leakage \citep{2011ApJ...730....5H,2011ApJ...741L..17Z}, the starburst galaxies Haro 11 \citep{2006A&A...448..513B,2011A&A...532A.107L}\footnote{The \citet{2006A&A...448..513B} detection was later disputed by \citet{2007ApJ...668..891G}, but the recent re-analysis by \citet{2011A&A...532A.107L} confirms the detection and places the leakage at a level of $f_\mathrm{esc}\approx3\%$} and Tol 1247-232  (Leitet et al. 2012, A\&A, submitted) are the only cases from which LyC photons have been directly observed. The question remains, however, to what extent a complete census of LyC leakage in the low-redshift Universe has yet been reached. Local LyC leakage candidates are typically selected on the basis of having strong emission lines of high excitation. While this guarantees that the target galaxies are producing ionizing photons in large numbers, this strong-line criterion will also bias the selection towards galaxies with  a high \ion{H}{i} content relative to the ionizing flux. In this case the \ion{H}{ii} regions will be radiation bounded, which actually prohibits LyC escape. Here, we explore an alternative strategy for selecting potential LyC leakers from the Sloan Digital Sky Survey\footnote{http://www.sdss.org/} (hereafter SDSS) in the hope of increasing the number of low-redshift galaxies with known LyC escape. Detailed studies of such local, spatially resolved targets may then shed light on the mechanisms responsible for LyC escape \citep[e.g.][]{2007MNRAS.382.1465H} in a way that would be staggeringly difficult for the LyC leakers identified at higher redshifts. Eventually, this should have important implications for our understanding of cosmic reionzation. 
 
A  flat $\Lambda$CDM cosmology with H$_0$=70 km s$^{-1}$ Mpc$^{-1}$, $\Omega$$_m$=0.3, and $\Omega$$_0$=0.7 was adopted throughout the article. All magnitudes are given in the AB system, and flux densities in f$_{\lambda}$.

\section{The escape fraction}
\label{section:escape}
Direct measurements of leaking LyC radiation from the reionization epoch galaxies are not feasible, since  the Ly$\alpha$ forest is too dense at these redshifts. Instead we are limited to the study of analogous objects at $z\lesssim 4-5$ \citep{2007MNRAS.382.1657K,2008MNRAS.387.1681I}. The amount of ionizing radiation leaking from galaxies can be quantified by the LyC escape fraction. For local galaxies the \textit{absolute escape fraction} ($f_{\mathrm{esc}}$) is mostly used,  defined as the ratio of the observed flux density of LyC photons ($f_{LyC,\mathrm{obs}}$) over the intrinsic LyC flux  density produced by the stars ($f_{LyC,\mathrm{int}}$). The parameter $f_{LyC,\mathrm{int}}$ is usually derived from the extinction corrected H$\alpha$ flux.

For distant galaxies the H$\alpha$ line is shifted out of the optical wavebands, and $f_{LyC,\mathrm{int}}$ is therefore difficult to access. Instead the \textit{relative escape fraction} ($f_{\mathrm{esc,rel}}$) is often used, defined as the ratio of the intrinsic Lyman break amplitude derived from SED models, ($f_{1500}/f_{LyC})_{\mathrm{int}}$, over the observed amplitude, ($f_{1500}/f_{LyC})_{\mathrm{obs}}$, corrected for the IGM optical depth. To compare the relative escape fraction of distant galaxies with the absolute escape fraction of local galaxies, the former needs to be corrected for the internal absorption by dust at 1500 {\AA}, $A_{1500}$:

\begin{eqnarray}
f_\mathrm{esc} = f_{\mathrm{esc,rel}}  \times 10^{-0.4 \cdot A_{1500}} = \\
\nonumber \centering \frac{(f_{1500}/f_{LyC})_\mathrm{int}}{(f_{1500}/f_{LyC})_\mathrm{obs}} \times exp(\tau_{LyC}^\mathrm{IGM}) \times 10^{-0.4 \cdot A_{1500}},  
\label{eqfesc}
\end{eqnarray}

where $\tau_{LyC}^\mathrm{IGM}$  is  the IGM optical depth for the LyC photon. $f_{LyC,\mathrm{obs}}$ is usually measured at rest wavelength $\approx$900 {\AA}, but for $z$ $\sim$ 1 galaxies the LyC has been sampled at shorter wavelengths even down to $\approx$700 {\AA} (\citealt{2003ApJ...598..878M,2007ApJ...668...62S,2010ApJ...723..241S,2009ApJ...692.1476C,2010ApJ...720..465B}). In the literature, the intrinsic amplitude has often been assumed to be ($f_{1500}/f_{LyC})_\mathrm{int}$=1-2.5 (3-7 in units of f$_{\nu}$) for a normal stellar population based on Starburst99 (e.g. \citealt{2007ApJ...668...62S,2009ApJ...692.1287I}). This agrees with predictions from our models \citep{2001A&A...375..814Z}. 

It has been suggested that the ionizing emissivity of galaxies is evolving with redshift (e.g. \citealt{2006MNRAS.371L...1I,2010ApJ...723..241S}). Many investigations at $z\approx3$ do indicate high escape fractions, both for single detections and stacked samples, while the low-$z$ universe shows a different picture. In the following we have made an effort to summarize and visualize the existing data in one single plot. To do this we have obtained the  observed ratios ($f_{1500}/f_{LyC})_\mathrm{obs}$ from the literature. Since the treatment of the IGM opacity varies between different authors, we have homogenized $\tau_{LyC}^\mathrm{IGM}$  using the prescription in \citet{2011ApJ...736...18N} around $z$ $\approx$ 3, and \citet{2010ApJ...723..241S} for $z$ $\approx$ 1. This means, for example, that a galaxy at z=3.2 will have $\tau_{LyC}^\mathrm{IGM}$=0.95 while one at z=1.3 will have $\tau_{LyC}^\mathrm{IGM}$=0.59. One should be aware that the dispersion in optical depth can be quite substantial for different sightlines \citep{2008MNRAS.387.1681I,2011ApJ...736...18N}, but here we can only apply the mean. We have also used a single value for the intrinsic Lyman break amplitude ($f_{1500}/f_{LyC})_{\mathrm{int}}$, even though this is known to be strongly dependent on  parameters such as the burst age and metallicity, and could in principle also be evolving with redshift.  Here, we have used the intrinsic ratio derived from Haro 11, which is the only LyC leaking galaxy yet where the SED model has been constrained by observations in multiple filters ranging  from the far-ultraviolet (FUV) to the near infrared (NIR), including the  H$\alpha$ emission line \citep{2007MNRAS.382.1465H}. The intrinsic ratio for Haro 11 was found to be ($f_{1500}/f_{LyC})_\mathrm{int}$=1.5 (Leitet et al. 2012, A\&A, submitted), somewhere in the middle of the range of values used in previous papers.  The value 1.5 is valid for a stellar population older than $\sim$20 Myr, using the  \citet{2001A&A...375..814Z} spectral synthesis model. For the galaxies where the LyC is sampled at shorter wavelengths, we have made a correction to the intrinsic ratio following \citet{2007ApJ...668...62S}. The internal absorption, E(B-V), was adopted as either the value reported by the author, or where no value was reported, as the mean value from similar galaxies at the same redshift. To derive the extinction at 1500 {\AA}, we have used the \citet{1997AJ....113..162C} extinction law where $A_{1500}$=10.33$\times$E(B-V).

\begin{figure}[t!]
\centering
 \resizebox{\hsize}{!}{\includegraphics{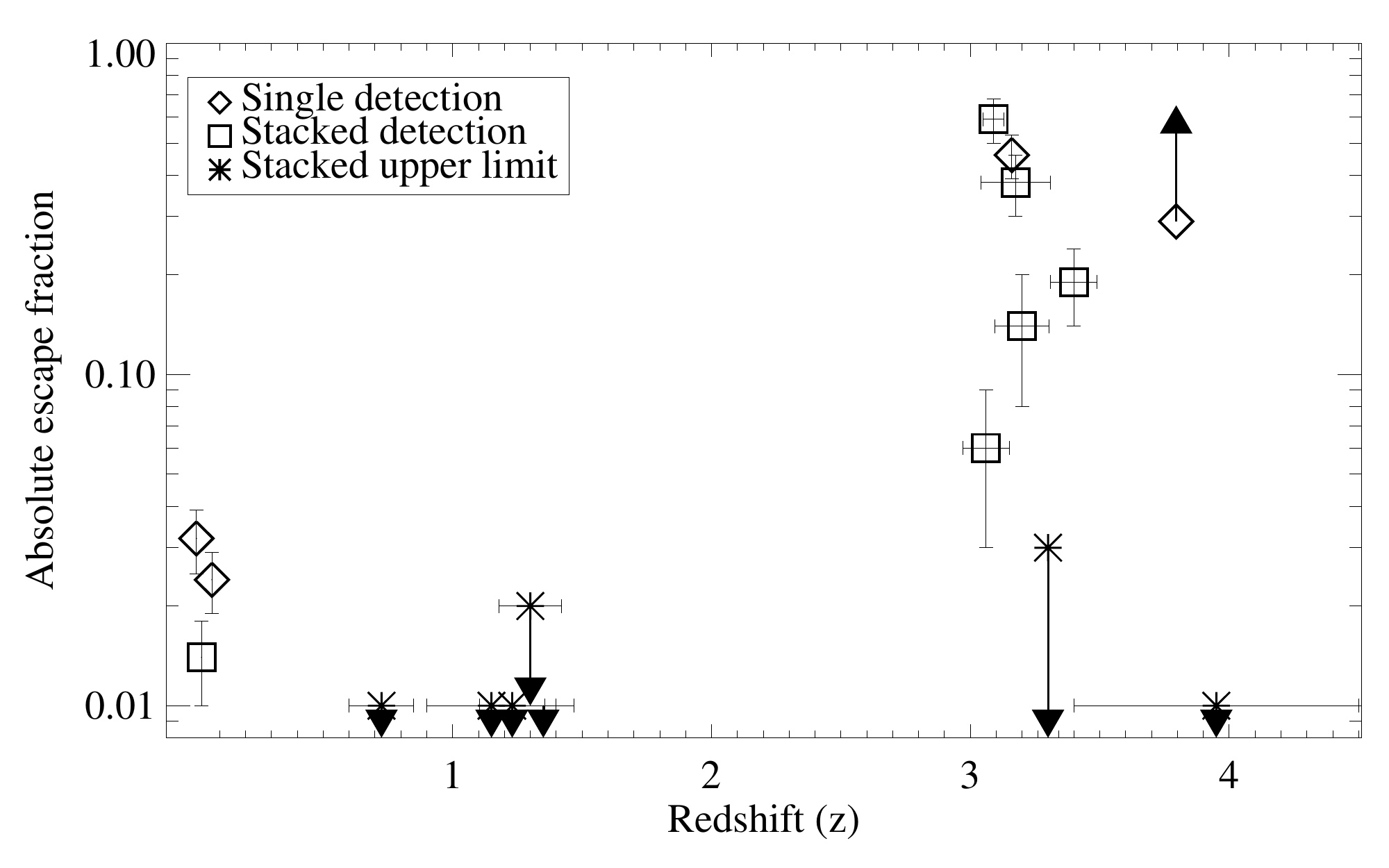}}
\caption{The absolute escape fraction, $f_{\mathrm{esc}}$, as a function of redshift. The data points are a mixture of single detections with errors, detections in stacked samples with errors and upper limits from stacked samples. The absolute escape fraction has been calculated from reported ($f_{1500}/f_{LyC})_\mathrm{obs}$ values, using a homogenized treatment of the IGM opacity and the intrinsic ratio ($f_{1500}/f_{LyC})_\mathrm{int}$, as described in the text.}
  \label{fesc}
\end{figure}

The escape fraction as a function of redshift is shown in Fig.~\ref{fesc}. The dispersion in the ($f_{1500}/f_{LyC})_{\mathrm{int}}$ model data has not been included in the error budget, only the observational uncertainties. The data are obtained from the following papers. At $z \approx$ 3 we use the stacked sample from \citet{2001ApJ...546..665S}, the stacked sample and the one confirmed single detection from \citet{2006ApJ...651..688S}, the stacked sample from \citet{2009ApJ...692.1287I}, the stacked sample and the z=3.795 single detection (with a lower limit dependent on the IGM opacity) from \citet{2010ApJ...725.1011V}, the stacked samples of Lyman break galaxies (LBGs) and \lya emitters (LAEs) from \citet{2011ApJ...736...18N}, and the stacked sample from \citet{2011ApJ...736...41B}. At $z$ $\approx$ 1 we use the stacked samples of \citet{2003ApJ...598..878M,2007ApJ...668...62S,2009ApJ...692.1476C,2010ApJ...723..241S} and \citet{2010ApJ...720..465B}. In the local universe we use the single detection from \citet{2011A&A...532A.107L}. In the local universe we have also included a new detection, for the galaxy Tol 1247-232, as well as the stacked spectrum of 9 FUSE galaxies (Leitet et al. 2012, A\&A, submitted). We have chosen not to include upper limits based on single galaxies.

From the limited information we have, the results suggest an overall lower  $f_\mathrm{esc}$ for the present day galaxies than for the galaxies at $z$ $\approx$ 3, but the relation is weakened with the recent results from  \citet{2010ApJ...725.1011V} and \citet{2011ApJ...736...41B} where only upper limits could be derived for their stacked samples. There is also a substantial risk that some of the $z$ $\approx$ 3 detections are actually low redshift interlopers \citep{2010MNRAS.404.1672V,2012ApJ...751...70V}.

Here, we argue that the few reports of LyC leaking galaxies at lower redshifts, and  the apparent decrease in $f_\mathrm{esc} $ at $z\lesssim 3$, may not only be the result of an evolving escape fraction. There are also several possible biases to consider. One such bias concerns at what wavelengths the Lyman continuum is sampled. All investigations at $z \sim$1 have been carried out in the rest wavelength range 700-850 {\AA}, while most investigations at $z \sim$3 and in the local universe sample around 900 {\AA}. Even a small amount of dust, allowing some LyC photons to escape at 900 {\AA}, would have an increasing effect with decreasing wavelength. The general importance of dust and how it may contribute to the detection rates over a large redshift range is further discussed in  Sect.~\ref{section:redshift}. Another possibility is that if there would still exist pop III stars at $z \sim$1, the nebular "Lyman bump" emission just below the Lyman limit \citep{2010MNRAS.401.1325I,2011MNRAS.411.2336I} would be missed.

Another important factor is the bias caused by selection effects. The main difference between surveys at $z \approx$3 and those performed in the local universe is the target selection. Most single detections  have been made at z$\sim$3 when narrowband imaging is utilized just below the Lyman limit in multi-object fields. The selection criteria for these objects are only weakly constrained by including LBGs and LAEs within a certain field and redshift range. On the contrary, the selection of galaxies observed in the local universe is heavily constrained by the assumed properties of LyC leakers. 

One urgent question now emerges - \textit{Are we really probing the local equivalent to the LyC-emitting galaxies
found at high z?}  Here technology is the limiting factor: multi-object fields cannot be observed, and we can only use spectroscopy (currently HST/COS). The only thing we can improve is the selection strategy of the targets. In the following sections we will present a novel method to select galaxies that are the most likely to have high escape fractions of ionizing photons, and take a closer look at a selection of $z \sim$ 0.03 LyC leaking candidates. 


\section{How to optimize the search for leaking galaxies}
\subsection{Basic conditions for LyC leakage}
\label{section:model}
In the local universe, dwarf galaxies involved in intense starbursts have been the main targets in searches for leaking galaxies. These galaxies have strong emission-line spectra with H$\alpha$  equivalent widths typically in the range \wha = 500 - 1000 \AA. In a galaxy with a significant leakage however, the nebular contribution would be much lower ($\propto$(1-f$_{esc}$) in the picket-fence model, see below), due to the lack of available gas to ionize. Although the colours would be blue, such a galaxy would have a 'dull' optical spectral appearence with weak emission lines. Therefore, they could naively be regarded as uninteresting and escape selection. If the leakage is due to a global ionization of a significant fraction of the \ion{H}{i} envelope (in the idealized case a truncated Str\"omgren sphere) the degree of ionization should be high. If, on the other hand the radiation is leaking through a cleared channel, the so-called 'picket-fence' model, it should be more normal. From what we know, these aspects of the expected observational properties of a LyC leaking galaxy have been neglected in previous searches. In this paper we describe a method to select galaxies with these properties. Our method finds further support from recent results in \citet{2011ApJ...736...18N} of a weaker Ly$\alpha$ emission from LyC leaking galaxies than non-leakers.

An additional selection criterion to be added is the age. The early stages of the reionization probably takes place in the approximate redshift range 9$<$z$<$11. This means that the starburst must be younger than the corresponding timespan, i.e. $\ltapprox$100 Myr. We thus have to limit our models discussed below, such that the maximum age of the young component is $\ltapprox$ 100 Myr.   

\subsection{Dust attenuation}
\label{section:dust}
An important issue concerns how to deal with the presence of dust in leaking galaxies. In the previous section we mentioned two basic ideas related to the conditions for LyC escape - the density bounded model (truncated Str\"omgren sphere) and the picket-fence model. Between these two extremes we find the more realistic density bounded ionization cone model where part of the light from the galaxy is escaping through the cone and part is absorbed in other directions. In the picket-fence model it is assumed that the radiation is leaking out through gas- and dustfree holes in the interstellar medium while the non-leaking regions are dusty. This condition may develop when gas outflows from the star forming regions open channels towards the halo. To create a density bounded cone model, the Lyman continuum flux needs to be intense enough to ionize all atomic hydrogen gas along the cone, allowing excess photons to disperse into the intergalactic medium. In this case the dust should be evenly distributed over the source, in the simplest case like a foreground screen. In the haloes of the galaxies the radiation density is much lower than in the starburst region. Under such conditions, even though the gas is ionized, small dust particles may absorb LyC photons without being destroyed \citep{1988ApJ...332..328B}. The critical radiation density is $\sim$ a few eV cm$^{-3}$. The \ha fluxes we derive from the target galaxies are typically 2-3$\times$10$^{33}$W, corresponding to a LyC flux of 2-3$\times$10$^{55}$eV s$^{-1}$. Assuming the produced photons to spread out isotropically, we obtain a radiation density of 1eV cm$^{-3}$ at a distance of $\sim$ 3 kpc from the source. Since this is below the critical radiation density for grain destruction, we would expect dust to survive at larger distances and contribute to attenuation of the light. A similar condition seems to be at hand in e.g. M82 \citep{1989ApJ...343L..13T}. The actual situation in leaking galaxies is probably somewhere in between the picket-fence and truncated Str\"omgren sphere models and in the following we will consider both of them to fully constrain the models.

\begin{figure}[t!]
\centering
\resizebox{\hsize}{!}{\includegraphics{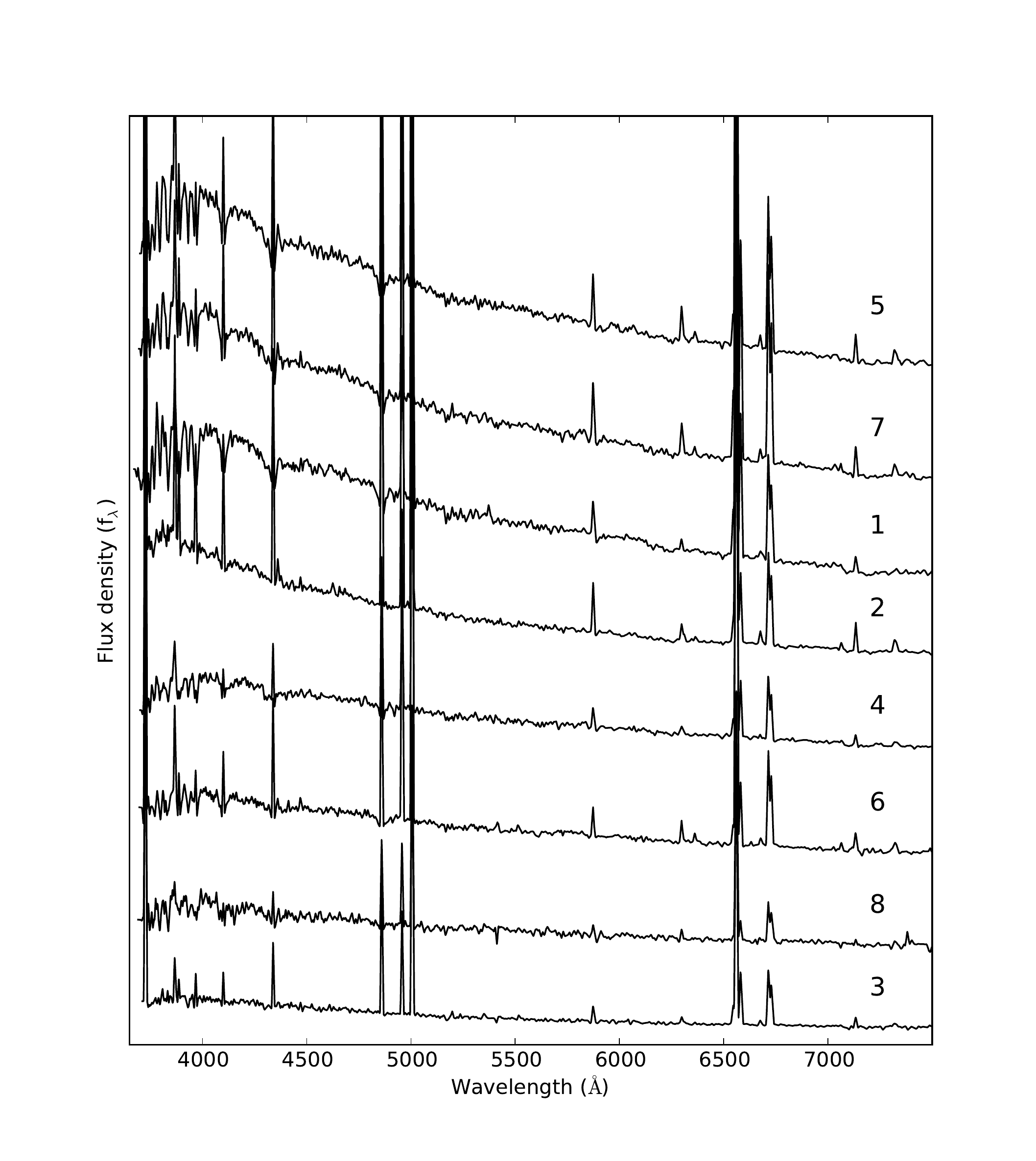}}
\caption{Optical spectra of the target galaxies, obtained from the 3" diameter fiber of the SDSS. The spectra have been shifted vertically with an arbitrary amount in order to make the individual spectra more clearly visible. The numbers refer to the running numbers in Table~\ref{table:basic}.}
  \label{spectra}
\end{figure}

\begin{table*}[htb!]
	\centering
	\caption{Basic data for the SDSS sample. Columns: 1) Running number 2) Name of the target galaxy 
	3) Right ascension 4) Declination 5) Redshift 6) Distance, obtained from NED 7) Absolute g magnitude based on SDSS {\sl model} magnitudes 8) Model u-g colour 
	9) Model g-r colour 10) Fiber u-g 11) Fiber g-r }
	\begin{tabular}{l l c c r r r r r r r}
	\multicolumn{11}{c}{}\\
	\hline \hline
        No & Name & RA ($\alpha$) & Dec ($\delta$)  &  $z$ & r & M$_g$ & u-g & g-r & (u-g)$_{f}$ & (g-r)$_{f}$   \\
	 &  & (J2000) & (J2000) &  &  Mpc & & & &  &    \\
	 (1) & (2) & (3) & (4) & (5) & (6) & (7) & (8) & (9) & (10) & (11)  \\
     	\hline
	1 & J010724.76+133209.2 & $01^h07^m25^s$ & $+13^d32^m09^s$  & 0.038 & 154 & -19.3 & 0.72 & 0.20 & 0.69 & 0.15 \\
     	2 & J024352.54-003708.4 & $02^h43^m53^s$  & $-00^d37^m08^s$ &  0.030 & 120 & -17.5 & 0.32 & -0.15 & 0.56 & -0.01   \\  
	3 & J025325.28-001356.6 & $02^h53^m25^s$  & $-00^d13^m57^s$ & 0.027 & 109 & -18.5 &  0.67 &  -0.17 & 0.33 & -0.09   \\  
     	4 & J075313.34+123749.1  & $07^h53^m13^s$ & $+12^d37^m49^s$ & 0.029 & 123 &  -17.6 & 0.74 & 0.12 & 0.69 & 0.11    \\  
	5 & J080754.64+141043.8 & $08^h07^m55^s$  & $+14^d10^m44^s$ & 0.029 & 122 & -19.3 & 0.81 & 0.29 & 0.59 & 0.13   \\  
	6 & J085642.06+123158.3 & $08^h56^m42^s$  & $+12^d31^m58^s$ & 0.029 & 127 & -18.3 &  0.82 & 0.24 & 0.66 & 0.21    \\  
       	7 & J100712.21+065735.6 & $10^h07^m12^s$  & $+06^d57^m36^s$ & 0.032 & 136 & -18.9  & 0.61 & 0.19 & 0.70 & 0.05   \\
      	8 & J101007.57+033130.5 & $10^h10^m08^s$  & $+03^d31^m31^s$  & 0.032 & 137 &  -18.3 & 0.65 & 0.25 & 0.71 & 0.17    \\     
    	\hline

	\end{tabular}
	\label{table:basic}
\end{table*}

\begin{table*}[htb!]
\centering
	\caption{Derived properties. Columns: 1) Running number 2) D4000n index 3) E(B-V) derived from the \hahb ratio after correction for underlying absorption 4) Fiber u-g colour, corrected for extinction assuming a foreground screen 5) Fiber g-r colour, corrected for extinction
	6) The [\ion{O}{iii}]$\lambda$5007/H${\beta}$ ratio 
	7) Oxygen abundance derived from the ON and ONS methods according to \citet{2010ApJ...720.1738P} 8) The exponential scale length in the \ha continuum 9) The \ha scale length
	10) Equivalent width of H$\alpha$  in emission derived from our measurements of the SDSS spectra (NEW) and as given in the SDSS data bank (SDSS) 11) The \ha equivalent width within a circular aperture of 10 arcseconds radius.}
	\begin{tabular}{l r r r r r r r r r r}
	\multicolumn{11}{c}{}\\
	\hline \hline
        No & D4000n & E(B-V)& (u-g)$_{f,c}$ & (g-r)$_{f,c}$ & [\ion{O}{iii}]/H${\beta}$ & 12+log(O/H) & $h_{H\alpha -cont}$ & $h_{H\alpha}$ & EW($H\alpha$) (\AA) & EW(\xha) (\AA) \\
	 &  &  &  &  & NEW/SDSS & ON/ONS & kpc & kpc  &  NEW/SDSS & Total  \\
         (1) & (2) & (3) & (4) & (5) & (6) & (7) & (8) & (9) & (10) & (11) \\ 
     	\hline
	1 & 1.01& 0.15 & 0.44 & -0.02 & 1.5/1.9 & 8.42/8.50 & 1.91& 1.41  & 67/67 & 81\\
	2 & 0.91 & 0.04 & 0.50 & -0.05 & 2.0/2.2 & 8.34/8.46 & - & -  & 166/180 & 66  \\
	3 &  0.82 & 0.16 & 0.11 & -0.28 & 3.8/4.0 & 10.2/8.7 & 1.56 & 1.14 & 230/228 & 170  \\
	4 & 0.95 & 0.22 & 0.32 & -0.15 & 3.1/3.4 & 8.11/8.28 & 1.36 & 0.59 & 146/146 & 78 \\
	5 & 0.95 & 0.13 & 0.37 &  -0.03 & 1.8/2.0 & 8.38/8.45 & 1.30 & 1.04 & 119/118 & 70 \\
	6 & 0.98 & 0.11 & 0.47 & 0.08 & 2.1/2.6 & 8.11/8.25 & 2.32 & 1.32 & 79/79 & 72 \\
	7 & 0.94 & 0.11 & 0.53 & -0.08  & 3.0/3.3 & 8.14/8.29 & 1.44 & 1.07 & 100/108 & 72  \\
	8 & 1.02 & 0.15 & 0.45 & -0.02 & 2.0/2.4 & 8.06/8.40  & 1.61 & 1.60 & 71/70 & 51  \\
   	\hline
	\\
	\end{tabular}
	\label{table:derived}
\end{table*}


\subsection{A tentative $z \sim$0.03 sample}
In order to have a closer look at galaxies with the desired properties in the local volume, we performed a first selection of leakage candidates in the SDSS at $z \sim$0.03 using the following selection criteria: 

\begin{itemize}
\item
0.025 $<$ $z$ $<$ 0.040 
\item
20 \AA $<$ \wha $<$ 200 \AA 
\item
2.5 $< $ \xhahb $ <$ 4
\item
\whb $>$ 10 \AA 
\item
\o5/\xhb ~$>$ 2 
\item
u-g $<$ 0.7 (prior to extinction corrections) 
\end{itemize}

In addition to this, AGNs were deselected according to the SDSS classification and the criterion that non-AGNs should have a Gaussian width of \ha of $\sigma_{H{\alpha}}$$< $5$\times$(z+1) \AA. Any remaining AGNs were later efficiently removed using the BPT diagnostics \citep{1981PASP...93....5B}. The lower redshift limit was chosen to make sure that the galaxies were distant enough for a large fraction of the light from the target to enter the 3" aperture of the SDSS spectrograph. The limiting magnitude for the SDSS spectroscopic survey is m$_r\sim$17.5. This corresponds to M$_r\sim$~-18 at $z$=0.03. To account for a reduced amount of nebular emission if the galaxy is leaking, we selected galaxies in the range 20\AA$\leq$EW(H$\alpha$)$\leq$200{\AA}. The lower limit was chosen in order to avoid too large uncertainties in the extinction corrections. The upper limit in combination with the colour criterion, u-g$<$0.7, strongly increases the probability of finding leaking galaxies in the sample, although it does not completely exclude non-leaking cases. We will argue below that a minimum requirement to fulfil the properties of three of our five reddest target galaxies is at least $f_{esc}=$20\%. This is valid under the assumption that the galaxies are dust-free. But as we will show below, if we take the dust attenuation into account, we arrive at a minimum absolute escape rate of $\sim$5\%. Galaxies with lower escape rates will be difficult to identify due to the uncertainties in the model predictions and we therefore chose this value, corresponding to \xwha $\sim$ 200\AA, as an upper limit. Our \ha observations described below, show that this criterion holds globally for all of the eight galaxies finally selected for observations. The extinction corrections are based on the \hahb ratio and at low EW:s the reliability of the correction is very sensitive to how one applies the correction for underlying absorption. To avoid galaxies with high extinction we set an upper limit to \hahb of 4 and to avoid problems with correction for underlying absorption we demand \whb to be larger than 10\AA. The \o5/\hb criterion guarantees that the excitation is high and thus that the light is dominated by massive shortlived stars. According to our models, this criterion is valid about 20 Myr after an abrupt shutdown of a starburst. The colour criterion is based on model predictions of a young stellar population mixed with an old host and includes a normal amount of interstellar extinction.  The galaxies were also selected to be visible from the ESO telescopes.  Of the $\sim$350000 objects catalogued within this redshift interval in the full SDSS data release 7, only 230 galaxies met the criteria ($<$0.1\%, see Sect. 5.2 for a discussion).  Next we calculated the extinction corrections based on the \hahb ratio, after correction for underlying absorption, using a method based on the D4000n break developed by \citet{2009IAUS..254P..44M}. The eight most promising LyC leaking candidates were then singled out for more detailed investigations. The SDSS spectra are displayed in Fig.~\ref{spectra}.

\begin{figure}[t!]
\centering
 \resizebox{\hsize}{!}{\includegraphics{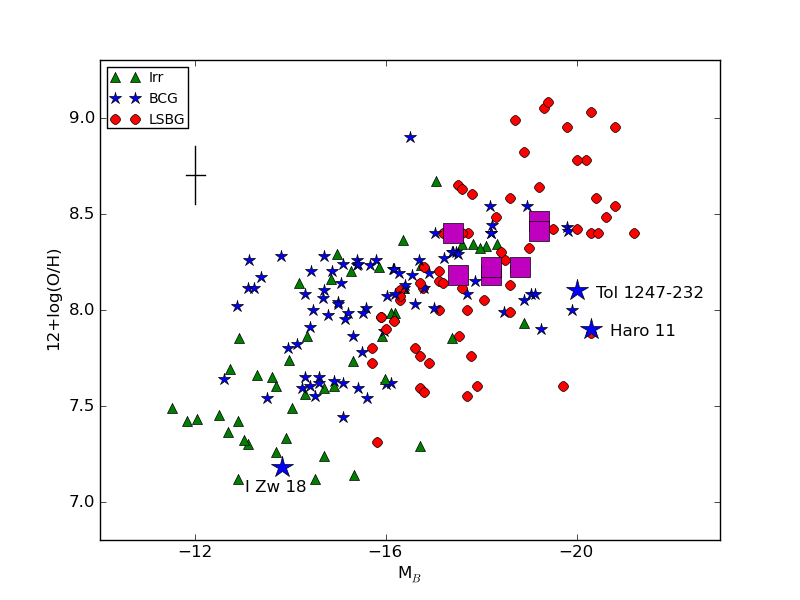}}
\caption{The luminosity-metallicity diagram for the $z \sim$0.03 galaxies. In the plot other local star forming galaxies of types Irregular, Blue Compact (BCG) and Low Surface Brightness (LSBG) are shown (Bergvall et al. 2012, in prep.). The eight potential LyC leakers, marked with large squares,  are distributed among the most luminous BCGs. Also shown are two known local LyC leakers, Haro 11 and Tol 1247-232. We have also indicated the much studied low metallicity galaxy I Zw 18. Typical 1$\sigma$ errors are indicated to the upper left.}
  \label{metal}
\end{figure}

The basic properties of the target galaxies are listed in Table~\ref{table:basic}, and the derived properties in Table~\ref{table:derived}. In column 7 of Table~\ref{table:derived} we give two alternative determinations of the oxygen abundances. Accurate abundances can be calculated if we know the electron temperature of the ionized gas. Normally the weak [\ion{O}{iii}]$\lambda$4363 line is used for this purpose. If the line is too noisy, semiempirical methods based on strong emission lines have frequently been used, the most popular ones being the R23 \citep{1979MNRAS.189...95P} and the N2 methods \citep{2002MNRAS.330...69D}. The reliability of the methods have to be tested from time to time as more accurate data become available. The strong line methods are therefore continuously being improved. A discussion of these issues with regard to SDSS data is found in the paper by \citet{2010JApA...31..121S}. Here we have used an extension of these methods as described by \citet{2010ApJ...720.1738P}. These are based on a combination of oxygen and nitrogen lines (ON method) and oxygen, nitrogen and sulphur lines (ONS method). One should take these abundances with a pinch of salt since, if the galaxies are leaking, the \ion{H}{ii} regions are not radiation bounded and the equations used to derive the oxygen abundances are not fully applicable. Anyway, the derived abundances  agree very well between the two methods and also between the eight galaxies except for galaxy no. 3. This galaxy is attributed a higher metallicity than the others, outside the metallicity range used for calibration of the ON and ONS methods. This probably explains the deviations between the predicted values for this galaxy. Typical abundances are 12+log(O/H) $\sim$ 8.3, corresponding to 40-50\% solar metallicity \citep[using the solar oxygen abundance as derived by][]{2004A&A...417..751A}.  This (40\% solar) is the metallicity we have adopted for our model comparisons in Fig. \ref{model} (see below). Comparing the eight potential LyC leakers with other local star forming galaxies, the galaxies of our sample are residing among the most luminous BCGs  in a metallicity-luminosity plot (see Fig.~\ref{metal}). But one should remember that the completeness limit at the typical redshift is M$_r\sim$~-18 so potential leakers may exist also at lower luminosities. The metallicities of the targets used in the diagram are mean metallicities as presented in Table~\ref{table:derived}. Data for the other galaxies are collected from various sources (Bergvall et al., in preparation). The known weak LyC leaker Haro 11 \citep{2011A&A...532A.107L}, often used as a local analogue of LBGs at $z \sim$3, is located to the extreme right in the distribution (see Fig.~\ref{metal}).

The low internal reddening estimated from the \hahb ratio indicates that the weak H$\alpha$ emission is not due to any abnormally low dust content for these galaxies.  Fig.~\ref{model} shows how a random sample of galaxies at $z \sim$0.03 with 20 \AA$\leq$EW(H$\alpha$)$\leq$200 {\AA} is distributed in the u-g/g-r diagram after correction for dust attenuation. The mean attenuation for the comparison sample in Fig.~\ref{model} is A(V)$\sim$0.45. Some guidance to how the colours relate to the spectral features are obtained from Fig.~\ref{filters}. We note that \ha is still in the $r$ window, i.e. not redshifted out. If the emission lines and the nebular continuum are weak, the colours essentially depend on the stellar continuum and this is the major trend we see in the evolutionary track with the highest escape fraction. The dramatic change in colour for this track is due to the change in age of the stellar population with the youngest population having the bluest colours. 

In Fig.~\ref{model} we also show the predicted colours of star forming galaxies fulfilling the \wha criterion, but with different amount of leakage, $b$-parameters and age. The models consist of mix a young and an old component, including a nebular component, and are described in more detail in Appendix A. In the figure we also see the positions of  our selected target galaxies, before and after corrections for dust attenuation, assuming a foreground screen. Unfortunately, the two confirmed leakers Haro11 and Tol 1247- 2321.4 are not observed in the Sloan survey so we cannot compare to their positions in the diagram. 

The total escape fraction for galaxies fulfilling our selection criteria can be estimated if we know the true dust attenuation correction in the UV. This correction is difficult to obtain so we will discuss two extreme cases - the picket-fence model and the homogeneous dusty foreground screen model. The actual conditions in the galaxies are probably intermediate between these cases. We split the escape fraction into two components: $f_{esc}$=$f_{esc,gas}\cdot f_{esc,dust}$, where $f_{esc,gas}$ is the fraction escaping through the neutral hydrogen gas and $f_{esc,dust}$ is the fraction remaining after passing through the dust. Let us first have a look at the picket-fence model. Here it is assumed that the leaking radiation is unattenuated so $f_{esc,dust}$=1. To allow for a comparison between observations and the models in Fig.~\ref{model} we should however correct the non-leaking regions for dust attenuation, using the \xha/\hb ratio. We do not know the exact proportions of the dust attenuated part and the leaking part. However, the model can help us to make a fair guess. Most of the targets have colours that, before correction, closely agree with the model shown with a dotted line. This model has a leakage of $f_{esc}$= 0.40. The solution is not unique. In fact, acceptable models can be found in the range 0.20$<f_{esc}<$0.60. But the 40\% model gives the best agreement with the observed colours and \wha of the targets. If we accept a deviation between model and observations of 30\% in \wha and 0.05 magnitudes in the colours, this model satisfies 80\% of the galaxies in the upper envelope. So we will assume that the escape rate is at least 40\%. If we add dust to the 60\% non-leaking part, the total colours will become redder. Therefore the leakage must be more than 40\%. By gradually increasing the fraction of leakage and correcting the non-leaking part for dust attenuation we find that the colours (the squares in the diagram) will experience a small shift towards bluer colours with the reddest colours corresponding to an absolute leakage of $f_{esc}$=$f_{esc,gas}\sim$50$\pm$10\%. 

Returning to Fig.~\ref{model}, we look at the second option - a dusty foreground screen. To correct the optical data for dust attenuation we used the observed \hahb ratios compared to the normal recombination ratio to derive the extinction coefficients. In Fig.~\ref{model} the corrected, dust-free colours are shown as triangles. The positions of the triangles can now be directly compared to the predictions from the dust-free model. We could of course make individual model fits to each of the target galaxies and then derive the predicted escape fractions from the observed \xwha. However, because the of the simple assumptions we make about the dust attenuation, which may vary with position and age of the stellar population \citep[e.g.][]{2000ApJ...539..718C}, we prefer to look at trends. Obviously most of the data points agree with a model that has leakage of $f_\mathrm{esc}\equiv f_\mathrm{esc,gas}\approx$ 0.95 and $f_\mathrm{esc,dust}$=1. Now we have to find the true value of $f_\mathrm{esc,dust}$ for a foreground screen with the derived $mean$ extinction coefficient.  Using the relations derived by  \citet{1994ApJ...429..582C} and \citet{2005ApJ...630..355C} we find that $<f_\mathrm{esc,dust}>$=0.1. Thus the absolute escape fraction as measured in front of the screen is $f_\mathrm{esc}\sim 0.95\cdot 0.1 \sim 0.1$. A more conservative value of the {\it minimum} escape rate of the sample galaxies is estimated to be $f_\mathrm{esc}\approx$0.05. Thus it is clear from this discussion that our target galaxies have properties that according to simple models allow for a substantial (5-50\%) leakage of LyC photons.

\begin{figure}[t!]
\centering
\resizebox{\hsize}{!}{\includegraphics{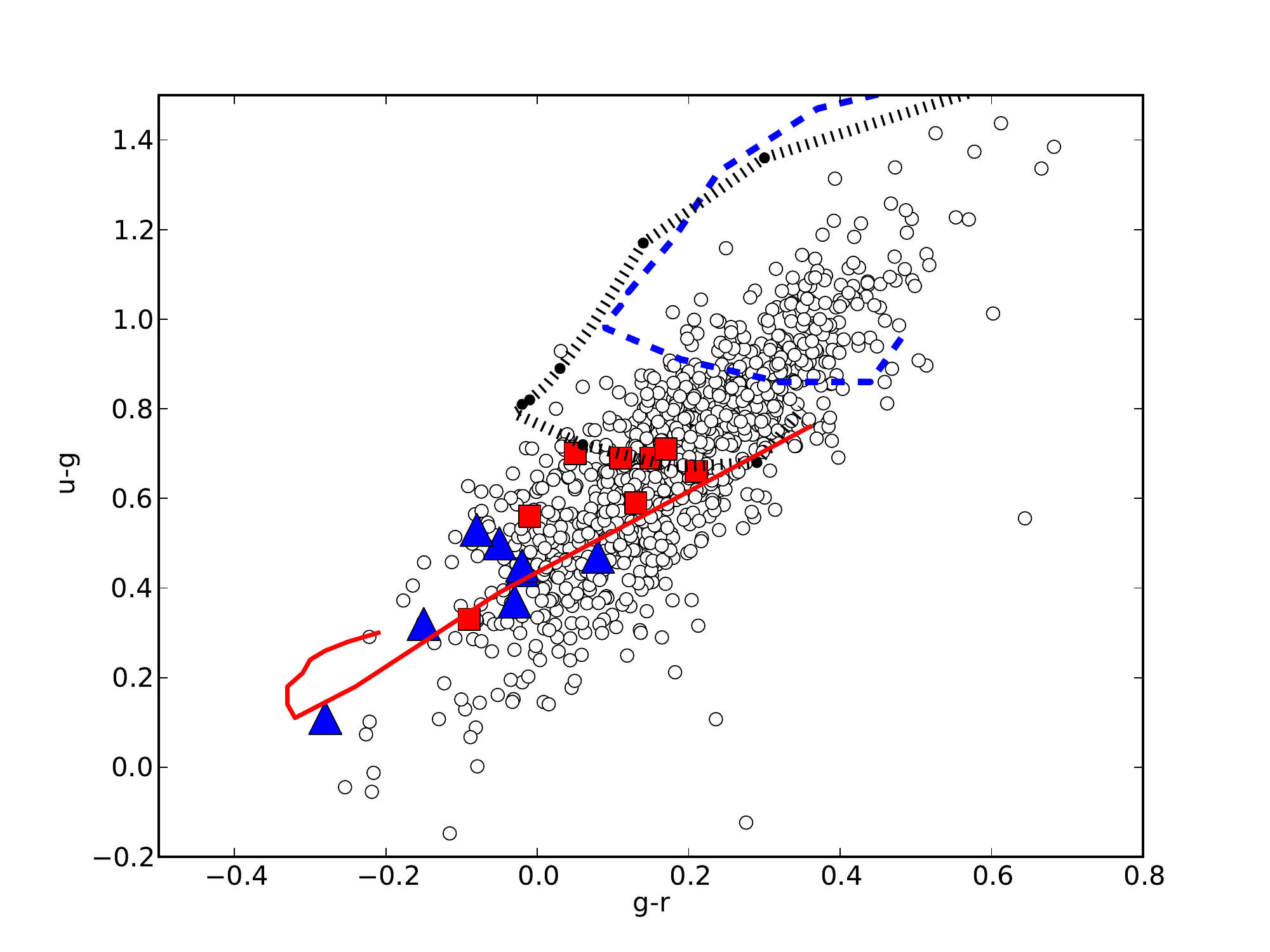}}
\caption{Comparison between model predictions and observed colours. The observed target colours are marked with squares. The triangles show the colours after correction for dust attenuation assuming a homogeneous foreground screen. The open circles show a random sample of 1000 star forming SDSS galaxies at z=0.03 with 20$\leq$EW(H$\alpha$)$\leq$200 {\AA}, also corrected for dust reddening. All corrections are based on the \ha/\hb ratio and are relevant for redshift z=0.03 and 40\% solar metallicity. Typical errors in the colours are 0.03-0.04 magnitudes but some of the outliers may be faint galaxies with large errors. Overlaid are tracks showing Zackrisson et al. (2001) model colours of galaxies with and without LyC leakage. Only parts of the evolutionary tracks where the EW criterion is fulfilled are shown. We look for cases which can be approximated by a mixture of a starburst with constant star formation rate and an old component. The duration of the starburst is 10 Myr. These tracks can be characterized by two parameters: 1) the birthrate parameter $b$, i.e. the ratio between present star formation rate and the mean past star formation rate, 2) $f_\mathrm{esc}$. The age of the old component was assumed to be 10 Gyr. The lines represent the following scenarios: \textbf{Hatched}: b=1, $f_\mathrm{esc}$=0 ; \textbf{Dotted}: b=2, $f_\mathrm{esc}$=0.4; \textbf{Solid}: b=30, $f_\mathrm{esc}$=0.95. The temporal evolution is from the top of the diagram. In the upper tracks we notice a kink on the blue side.  This marks the end of the starburst. The nearly horizontal part of the two upper tracks correspond to the post-starburst phase.}
  \label{model}
\end{figure}

\begin{figure}[t!]
\centering
 \resizebox{\hsize}{!}{\includegraphics{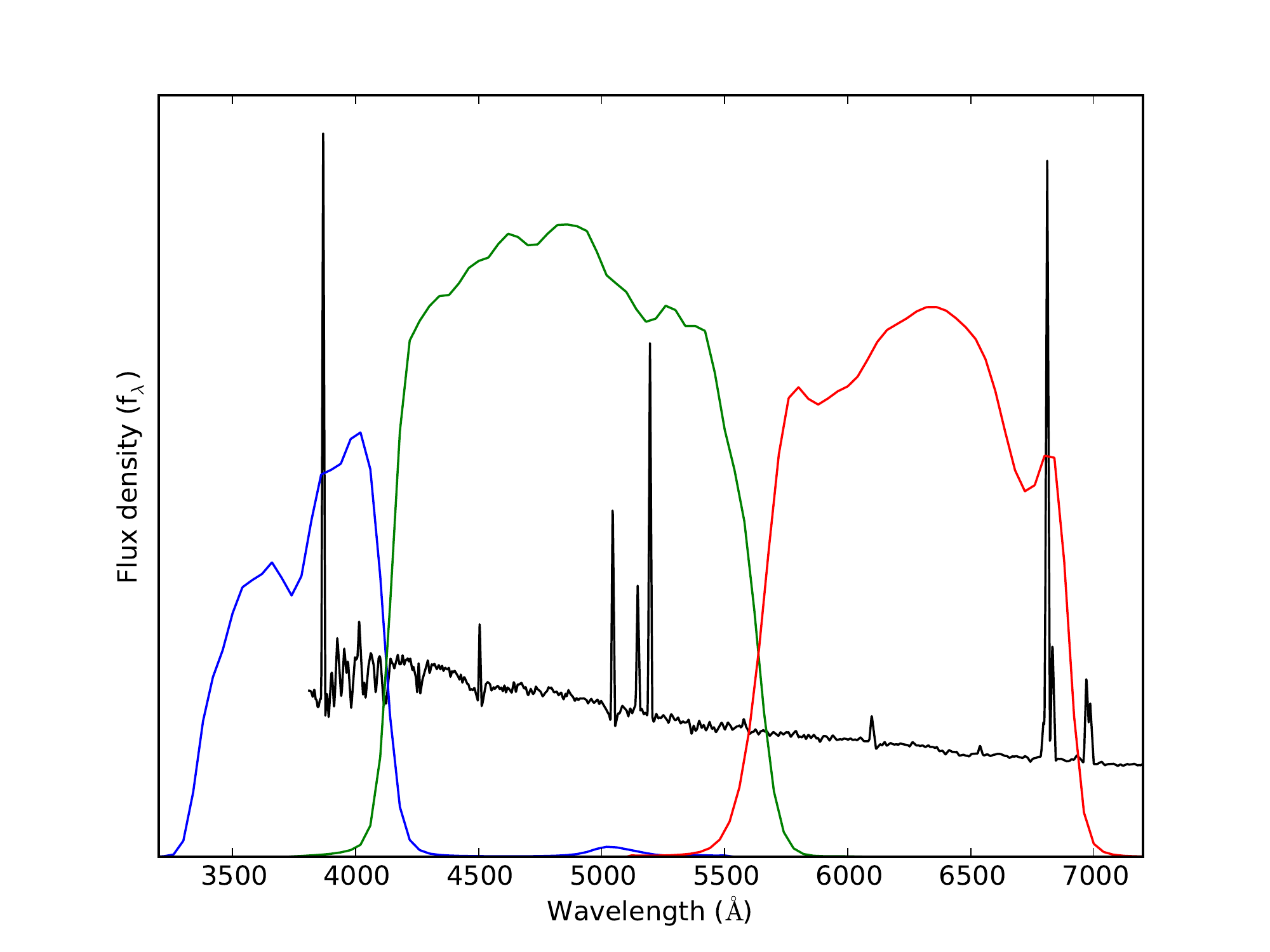}}
\caption{The spectrum of target galaxy No. 1 with the SDSS $ugr$ filters overlaid.}
  \label{filters}
\end{figure}



\subsection{Optical imaging}
\subsubsection{Observations and data reduction}
\label{section:obs}
The $z \sim$0.03 galaxies were imaged with the EMMIR and SUSI2 at the ESO New Technology Telescope (NTT) at La Silla, Chile, December 13-15, 2007. The three nights were dark and had good photometric conditions. The average seeing was 1.2\arcsec - 1.5\arcsec, with occasional periods of $\sim$ 2\arcsec seeing. H$\alpha$  images were obtained with EMMIR. We used the narrowband filters $\#$599 and $\#$600 (both with FWHM $\sim$ 71 \AA), having central wavelengths corresponding to the redshifted \ha line of the galaxy.  The H$\alpha$ offline filter 
($\#$768) with central wavelength 6014 \AA ~was selected to only include the stellar continuum not too far from the emission line. Images in the Cousins $B$ band were obtained with SUSI2. The instrument set-up and integration times can be seen in Table~\ref{table:obs}. The data were reduced  with the NOAO/IRAF package, using standard procedures including bias subtraction, flat-field correction, spectrophotometric standard stars, and sky subtraction. The H$\alpha$ images were produced from the online images after subtracting the offline images, scaled to make the stars in the field disappear after subtraction. A comparison with the SDSS \wha was carried out as a final check. The total \wha within a radius of 10 arcseconds is given in Table 2. These values are all within our search criteria. In Table~\ref{table:obs} are also given the exponential scale lengths in \ha and the \xha-continuum. As can be seen from Fig.~\ref{ha}, the \xha-emission is close to spherical, or the morphology is so distorted that it is difficult to define a symmetry plane. When we derived the scale lengths we therefore did not correct for inclination effects. We integrated radially and azimuthally from a position corresponding to the brightest part of the \ha image. In the derivation of the scale length, the central regions were excluded and only the exponential part of the images outside the very centre was used.

\begin{figure*}[t!]
\centering
\includegraphics[width=18.5cm]{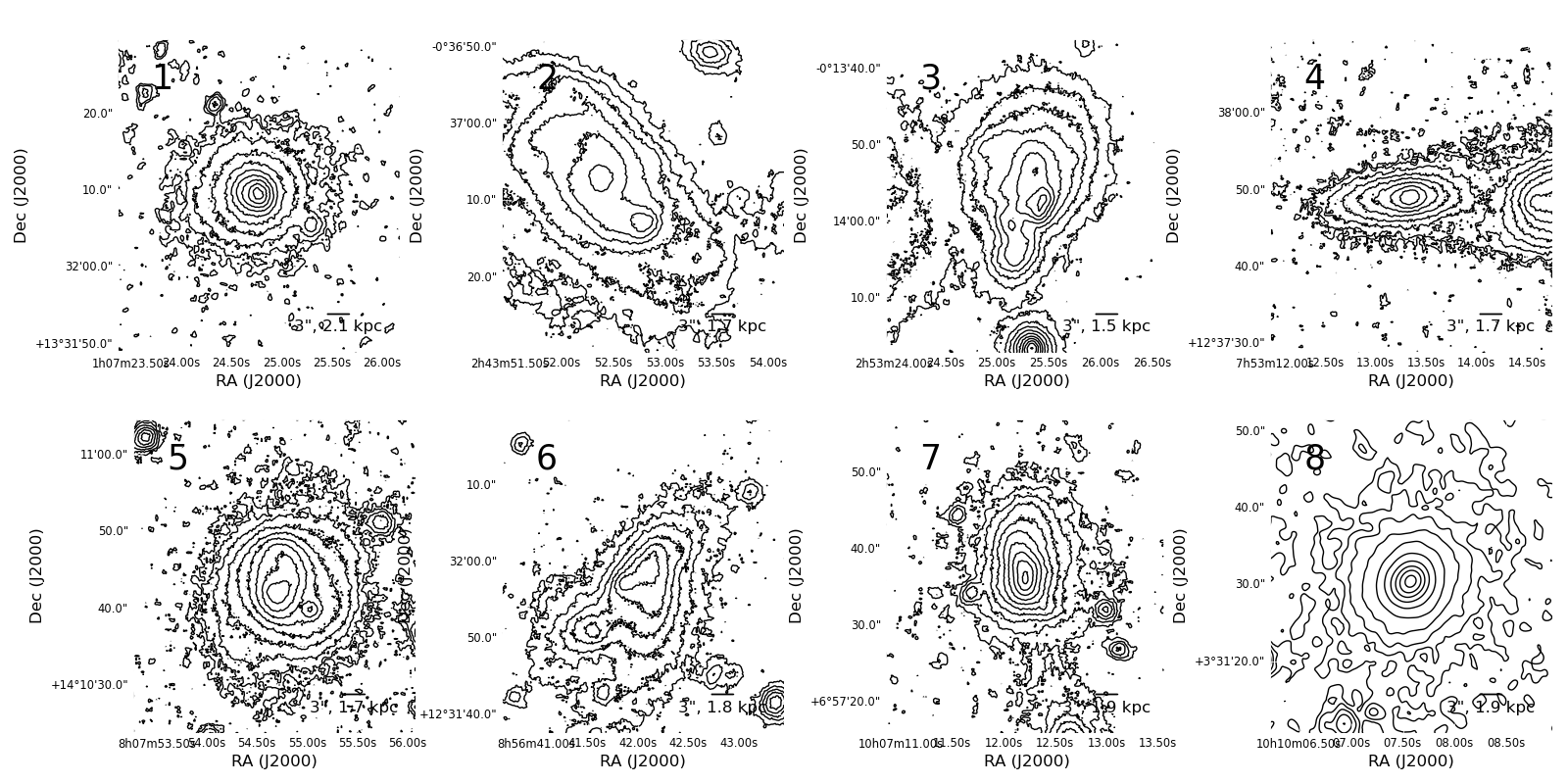}
\caption{Images in the \ha continuum of the target galaxies obtained with EMMI/NTT at ESO. East is left and north is up. The contour levels differ by a factor of 2. The numbers refer to the running numbers in Table~\ref{table:basic}.}
\label{hacont}
\end{figure*}

\begin{figure*}[t!]
\centering
\includegraphics[width=18.5cm]{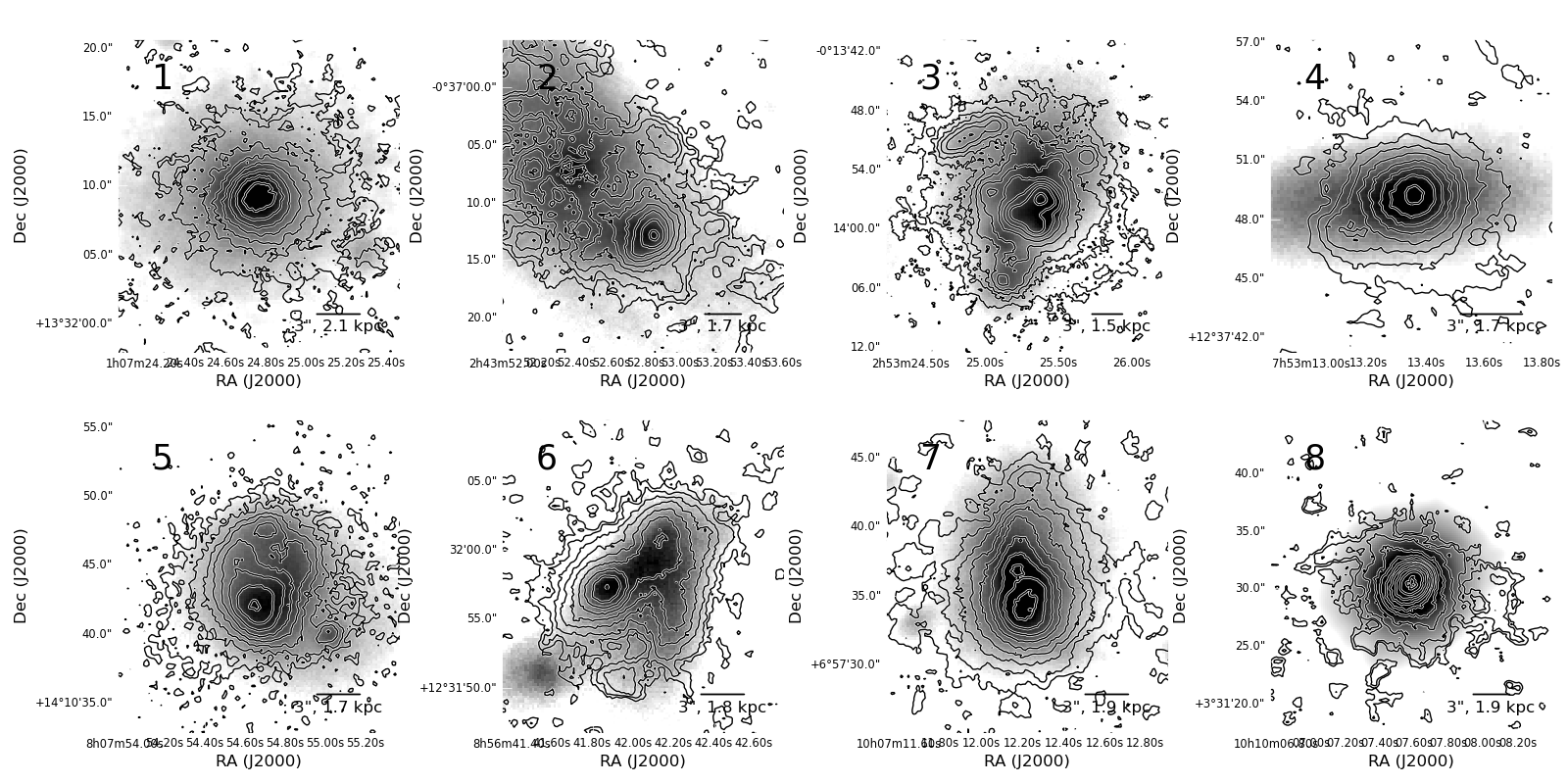}
\caption{Images of the target galaxies. East is left and north is up. The contours show the H$\alpha$ emission, while the inverted grey scale background shows the continuum from the H$\alpha$ off-line spectral region. The numbers refer to the running numbers in Table~\ref{table:basic}.}
\label{ha}
\end{figure*}

\begin{figure*}[t!]
\centering
\includegraphics[width=18.5cm]{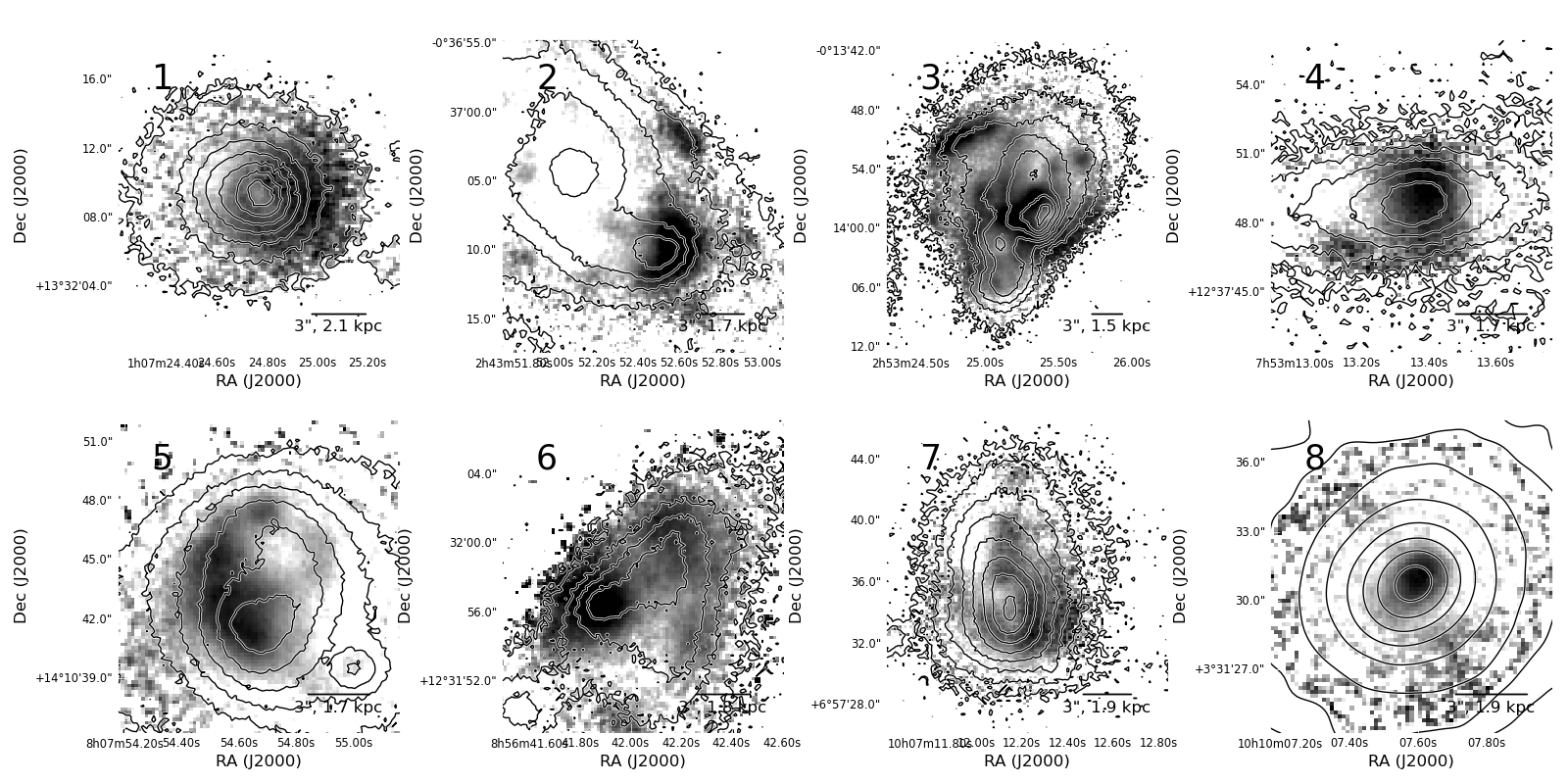}
\caption{Here is shown the \ha equivalent width in grey scale with the \ha continuum background shown in the isophotal map. The flux levels are arbitrary and are intended to give a  good overview of the stellar+free-free emission component. The numbers refer to the running numbers in Table~\ref{table:basic}.}
\label{wha}
\end{figure*}

\begin{table*}[htb!]
\centering
	\caption{The NTT $z \sim$0.03 sample observational details. Column: 1) Running number 2) Name of the target galaxy  
	3) H$\alpha$ online instrument/filter (exposure time in seconds)  4) H$\alpha$ offline instrument/filter (exposure time in seconds) 	5) the B images instrument/filter (exposure time in seconds).}
	\begin{tabular}{l l c c l}
	\multicolumn{5}{c}{}\\
	\hline \hline
        No & Object & H$\alpha$ online & H$\alpha$ offline & B filter \\
         (1) & (2) & (3) & (4) & (5) \\
     	\hline
	1 & J010724.76+133209.2 & EMMIR/$\#$600 (3100) & EMMIR/$\#$768 (1770)  & SUSI2/$\#$811 (800) \\ 
     	2 & J024352.54-003708.4 & EMMIR/$\#$599 (3100) & EMMIR/$\#$768 (1770) &  SUSI2/$\#$811 (1200) \\  
	3 & J025325.28-001356.6 &  EMMIR/$\#$599 (3100) & EMMIR/$\#$768 (1770) &  SUSI2/$\#$811 (1200) \\  
     	4 & J075313.34+123749.1  & EMMIR/$\#$599 (2190) & EMMIR/$\#$768 (930) & SUSI2/$\#$811 (960) \\  
	5 & J080754.64+141043.8 &  EMMIR/$\#$599 (2190) & EMMIR/$\#$768 (930) & SUSI2/$\#$811 (960) \\  
	6 & J085642.06+123158.3 & EMMIR/$\#$599  (2100) & EMMIR/$\#$768 (900) & SUSI2/$\#$811 (960) \\  
       	7 & J100712.21+065735.6 & EMMIR/$\#$599 (4380) & EMMIR/$\#$768 (1860) & SUSI2/$\#$811 (960) \\  
      	8 & J101007.57+033130.5 & EMMIR/$\#$599 (2190)  & EMMIR/$\#$768 (930) &  SUSI2/$\#$811 (960) \\   
		
    	\hline
	\\
	\end{tabular}
	\label{table:obs}
\end{table*}

 \begin{figure*}
  \centering
  \begin{minipage}[c]{0.21\linewidth}
     \centering{\includegraphics[width=4.35cm]{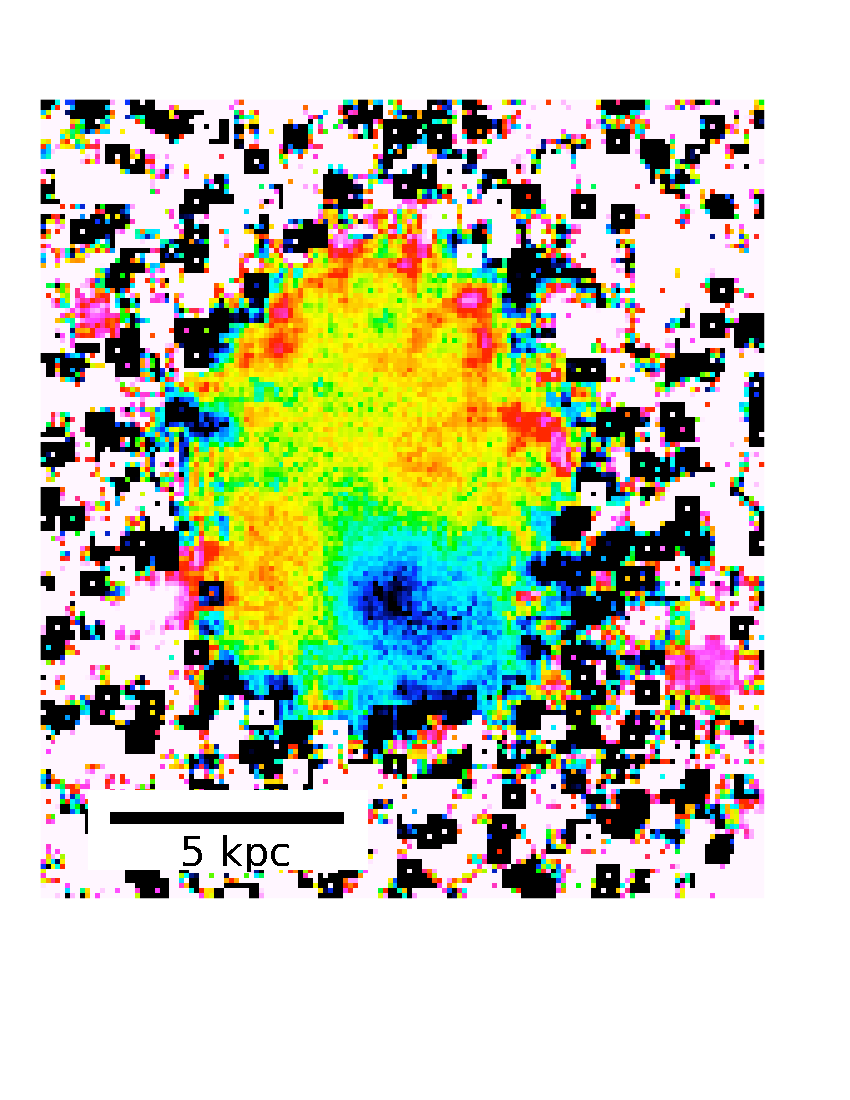}}
  \end{minipage}%
  \begin{minipage}[c]{0.24\linewidth}
     \centering{\includegraphics[width=4.35cm]{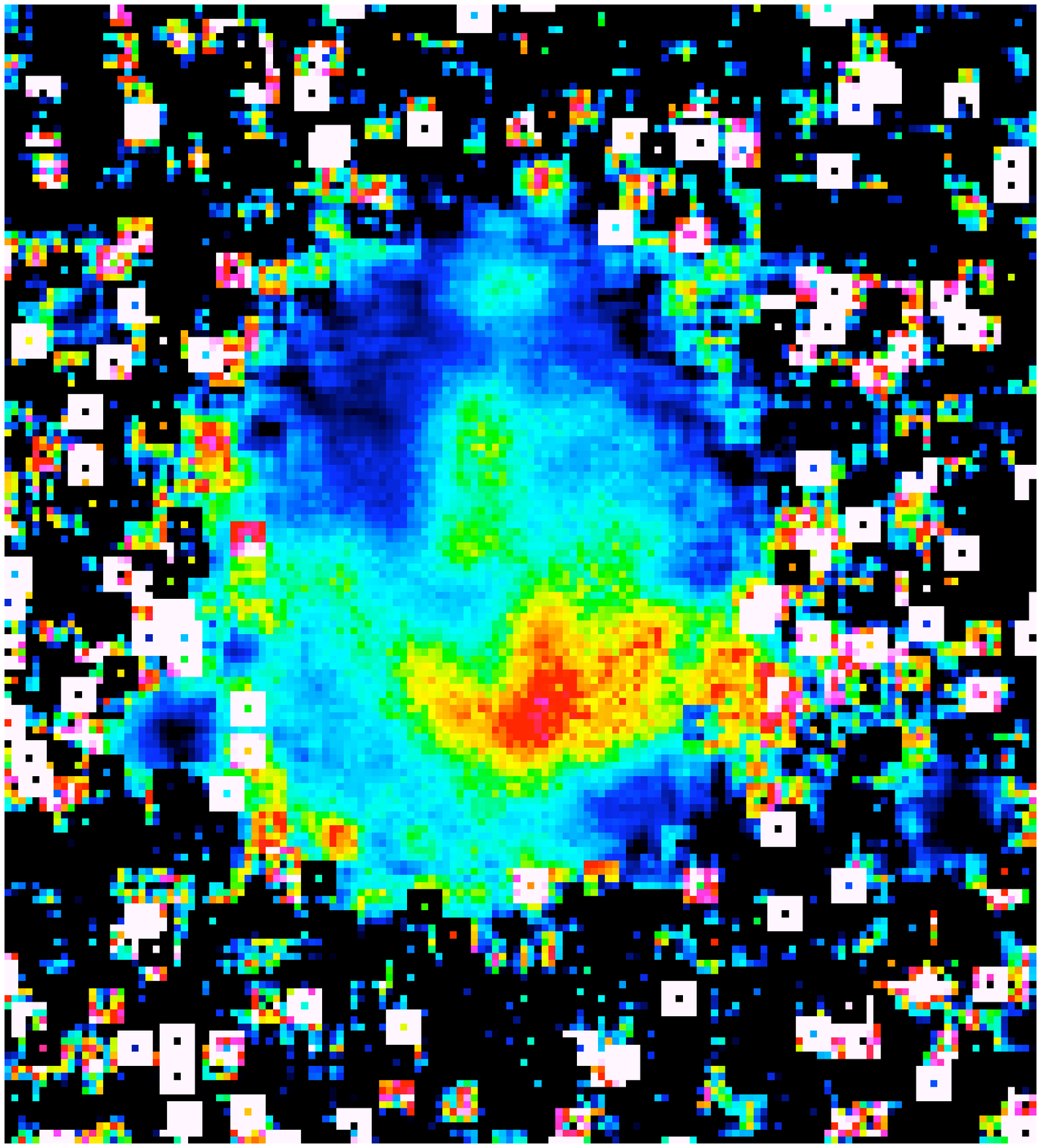}}
  \end{minipage}
  \begin{minipage}[c]{0.18\linewidth}
     \centering{\includegraphics[width=3.5cm]{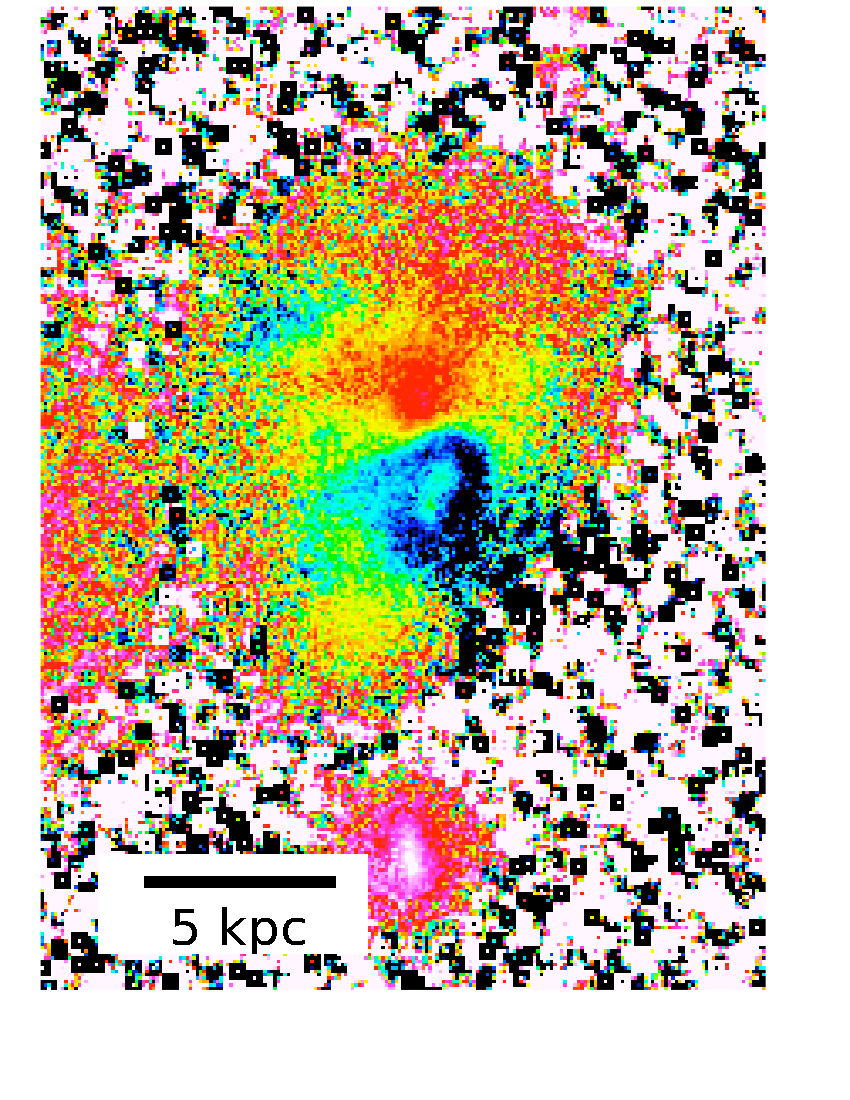}}
  \end{minipage}%
  \begin{minipage}[c]{0.18\linewidth}
     \centering{\includegraphics[width=3.5cm]{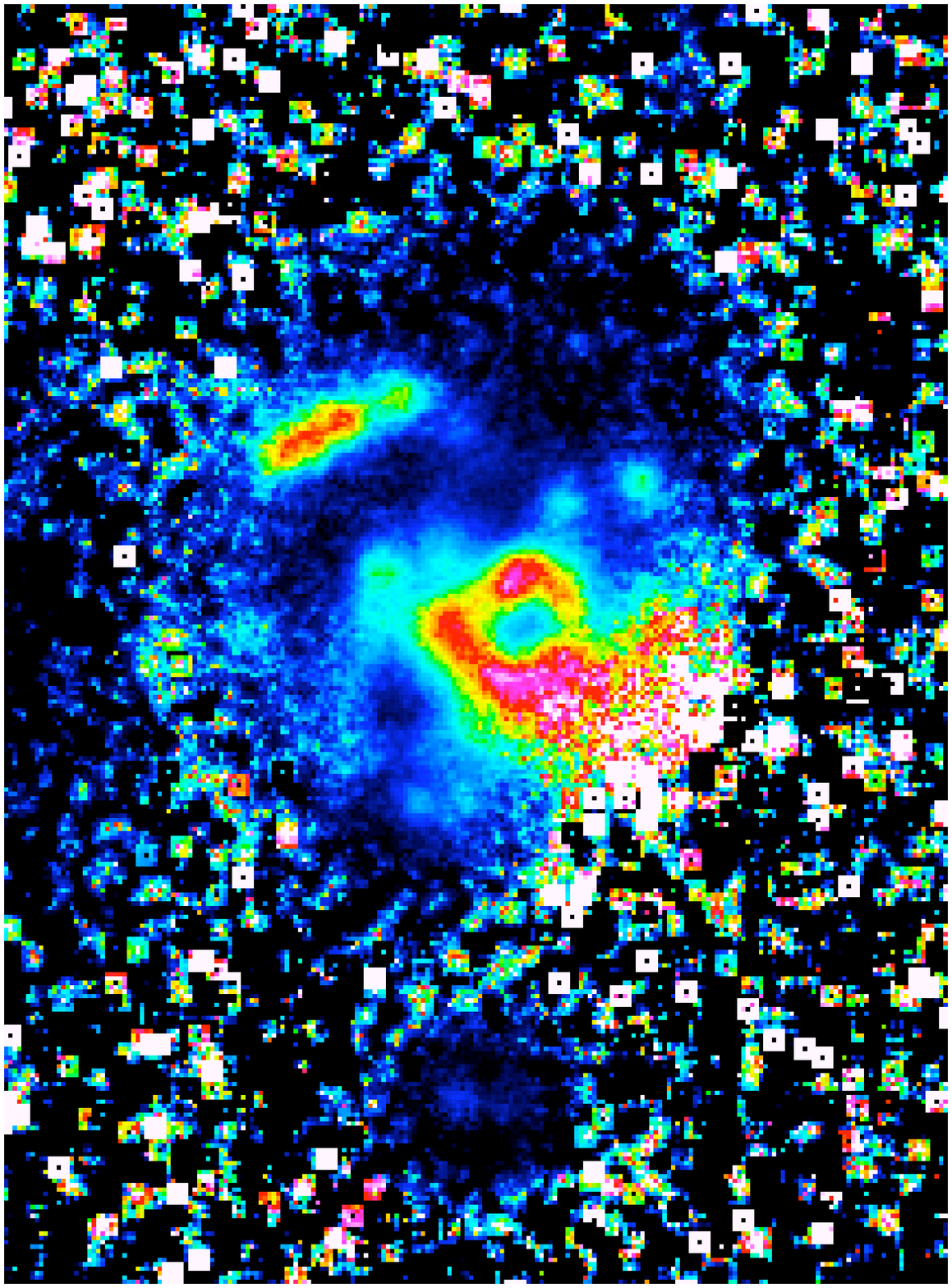}}
  \end{minipage}
     \caption{ The target galaxies no. 7, J100712.21+065735.6 (left pair) and no. 3, J025325.28-001356.6 (right pair). The left image in each pair shows the colour distribution based on the ratio between the fluxes in the B and  \ha-continua,  index B-EMMI768. Blue regions map young stars and red regions map old stars. The image on the right in each pair shows a map of  \xwha, ranging between 0 and 200 \AA ~Êin the galaxy on the left and  between 0 and 600 \AA ~Êin the galaxy to the right (white regions are higher values). Blue regions have low equivalent widths, red have high equivalent widths. The images have been selectively (depending on the local S/N) filtered with a median boxcar filter, using the MIDAS task {\it filter/smooth}.  North is up, east is to the left. The fieldsizes are 24"x26" and  36"x48" respectively.}
        \label{br_wha}
  \end{figure*}

\section{Results}

\subsection{Morphologies}
The target galaxies are shown in  Fig.~\ref{hacont}, ~\ref{ha} and ~\ref{wha}. The first figure shows the isodensity contours of the galaxies in the \ha continuum band. The second figure shows the \ha emission in isophotal contours with \ha continuum in greyscale. The third figure shows the \wha greyscale map overlaid on the isodensity contours of Fig.~\ref{hacont}. From a look at Fig.~\ref{hacont} we can see that it appears that there are two types of galaxies. One is dominated by a nearly spherical component (no. 1 \& 8). Others show a disturbed morphology with a a few galaxies (no. 4, 5 \& 7) displaying a tadpole structure \citep{1996AJ....112..359V} in \xha. Although our galaxies are fairly bright, albeit sub-L$^*$, this is uncommon among normal galaxies at these luminosities. The general impression is that most of the target galaxies appear to be dynamically disturbed, either with clear signs of interaction or clearly non-symmetric light distribution. Two galaxies, no. 1 and 8, at first appear to be in equilibrium, but also in these cases the \ha maps overlaid with the continuum maps reveal clear asymmetric structures.

\subsection{Young stars and ionized gas - implications for LyC leakage}

How can we use our observational data to better understand the conditions for leakage? We require from the global data that \wha should be low while at the same time the stellar population should be young. But how can we find a stronger support for leakage locally? We show two examples of our targets in Fig~\ref{br_wha}. The figure shows maps of a colour index and \xwha. The index is defined by the ratio of the flux in the B band and the \ha continuum flux as measured in the EMMI filter $\#$768 at the effective rest wavelength 6026 {\AA}, which at the redshift of our galaxies corresponds to a rest wavelength of 5850 \AA. We call this index B-EMMI768. It is a strong indicator of young star forming regions. In both examples shown in the figure, we see that the young population is forming a 'channel' from the centre out to the very limit of the optical image. The conditions for leakage will be higher the closer that young stars are to the halo region. If the gas is ionized by stars at the edge of the main body (defined by the noise limit of the optical image), the observed \wha should be low. This is what we observe in galaxy no. 7 on the left. We can derive a rough estimate of the mean density in the ionized gas from

\begin{equation}
\mu_{H\alpha}(R)  = \frac{\alpha^{H\beta}_{eff}h \nu_{H\beta} r_{\alpha\beta} n_{p,0} n_{e,0}}{4\pi} \int_{-\infty}^{\infty}  f(r[l,R]) e^{-2r(l,R)/h_{HII}} dl,
\end{equation}

where $R$ is the projected distance from the centre (sometimes called the {\sl impact parameter}),  $l$ is the distance in the tangential direction along the line of sight  and $r=\sqrt(l^2+R^2)$ is the radial distance. We assume a spherical distribution of the ionized gas and an exponential density decrease where $h_{HII}$ is the scale length in {\sl mean} density of the ionized gas. $\mu_{H\alpha}$ is the surface brightness in \xha. $ \alpha^{H\beta}_{eff}$ is the effective recombination coefficient of \hb (3.034 10$^{-14}cm^{3}s^{-1}$) and $r_{\alpha\beta}$ is the emission line ratio between \ha and \hb (2.85), both obtained from \citet{1987MNRAS.224..801H} assuming an electron temperature of $T_e$=10$^4$K and a density of 100 cm$^{-3}$. Both these parameters are only weakly dependent on densities at low density and can be applied here even if we find much lower densities.  $h$ is the Planck constant, $\nu_{H\beta}$ is the frequency of the \hb emission line, $n_{p,0}$ and $n_{e,0}$ are the central proton and electron number densities of the ionized gas in cm$^{-3}$ ($n_{e,0}\sim n_{p,0}$ [$\equiv n_0$] is assumed here), $f(r)$ is the density dependent filling factor, i.e. the fraction of space that is filled with ionized gas, assuming a bimodal density distribution. 

From this we obtain

\begin{equation}
n_0 = \left [ \frac{4\pi\mu_{H\alpha}} {\alpha^{H\beta}_{eff}h \nu_{H\beta} r_{\alpha\beta}\int  f(r) e^{-2r(l,R)/h_{HII}} dl} \right ]^{1/2}   
\end{equation}

and $n=n_0e^{-R/h_{HII}}$ at $l=0$. To proceed, we need to know the scale length of the ionized gas and how the filling factor varies with distance from the centre. Here we will assume that the mean gas density follows the luminosity distribution in the \ha continuum. From Table \ref{table:derived} we see that the scale length in \ha is shorter than in the \xha-continuum which is as expected since the \ha emissivity goes as $n^2$. The calculations are not very sensitive to the scale length we choose but more to the change of filling factor with density. Determinations of the filling factor in the Milky Way and the Andromeda galaxy \citep{2004Ap&SS.289..207B,2006AN....327...82B} show a nearly linear dependence of the filling factor with the reciprocal electron density, $f \propto n_e^{-1}$.  The filling factor thus increases with decreasing density and in the diffuse ionized gas (DIG) it is fairly high. In regions of active star formation, it may be more porous due to the winds form the star forming regions \citep{1997ApJ...491..561M,2004A&A...425..899D}.  Here we will assume that the filling factor follows this behaviour and increases with distance from centre. For simplicity we will assume that the filling factor is constant along the line of sight towards the faint blue regions since the distance from the centre, and thus the filling factor, is almost the same along a scale length distance from the densest region in the tangential direction. We will have a look at two extreme values of the filling factor, $f$=0.02 and $f$=0.2, defining a range typically found in the DIG \citep[e.g.][]{2003ApJS..148..383M,2004AJ....128.2758M,2008A&A...490..179B}. We will assume an electron temperature of 10$^4$K and that the scale length of the mean density of ionized gas is the same as for the \ha continuum, $h_{HII}$ =1.5 kpc. Using ~$\mu_{H\alpha}$ = 3.8$\cdot$10$^{-6}$erg s$^{-1}$ cm$^{-2}$ sr$^{-1}$as measured towards the optical edge of the blue region of galaxy No.  7, we obtain a mean density of the ionized gas at about 4 kpc fom the centre of $<n_e> \sim$ 0.01-0.04 cm$^{-3}$. This is about 2 dex lower than the densities in the largest H II regions in other galaxies, having sizes of 100-200 pc or more \citep[e.g.][]{2001AJ....122.3017S} and a few times lower than the DIG in the halo regions in normal disk galaxies \citep[e.g.][]{2001ApJ...551...57C,2003ApJS..148..383M}. The low density we find is more similar to that of the outflow regions of starburst galaxies, e.g. M82  \citep{2007A&A...467..979B}. Although the uncertainties in our derived numbers may be significant, the very low values lends support to favourable conditions for leakage.

We can also imagine another scenario where the ionizing radiation is not produced in the outskirts. Maybe there is a more centrally located ionization source that is so powerful that it can ionize the medium all the way into the halo in a certain direction. If the scale length of the ionized gas is larger than that of the stars, \wha can (but need not) increase as we approach the halo and we can observe extremely high values at faint levels. We see this happen in the second example, galaxy no. 3, to the right in Fig~\ref{br_wha}. In this transition region we find extremely high values of  \xwha, about 1000 \AA.  This could indicate that the gas is ionized far outside the region where the ionizing radiation is formed, in the low density halo. Within the ionized medium there may be low density channels through which the radiation easily can escape. Now we will show below that the scale length in \ha in fact is systematically smaller than the stellar scale length so how can we explain the observations? If the galaxy is subject to ram pressure stripping, the stellar disk may be truncated in the direction of the movement through the intergalactic medium  \citep{2006AJ....131..716K}. We will discuss ram pressure stripping below. If this is the case in galaxy no. 3, the rapid increase in \wha is understood. The most important conclusion here however is that the ionizing stars in the two cases seem to form a 'bridge'  between the main body and the halo that would favour leakage. 

\begin{figure}[t!]
\centering
\resizebox{\hsize}{!}{\includegraphics{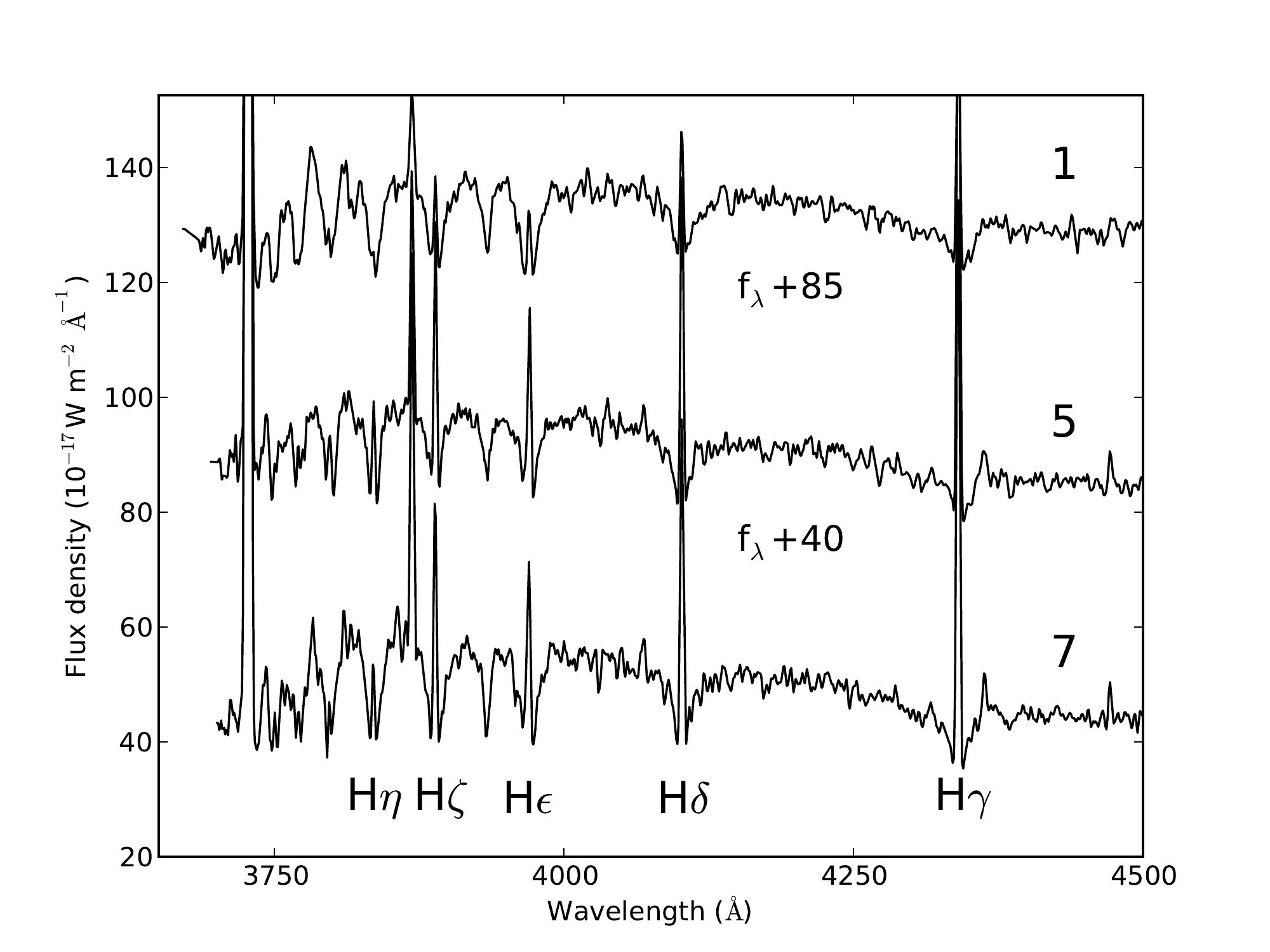}}
\caption{Spectra of the three galaxies showing post-starburst spectra. The numbers to the right refer to the running identification numbers given in the Tables. The strong Balmer lines in absorption are indicated.}
\label{postspec}
\end{figure}

Not all galaxies in our sample show a strong off-center star formation. This could partly be due to projection effects but also because we may be dealing with different types of galaxies, where the leakage becomes possible for different reasons. In two cases (galaxies 1 \& 8) the galaxies have very regular structures, deviating for the other non-equilibrium cases  (see Fig~\ref{hacont}).  Interestingly, these two galaxies are similar to the "dominant central objects" of the Lyman-break analogues defined as  LyC leakage candidates by \citet{2011ApJ...730....5H}. One of these galaxies, No. 1, shows a young post-starburst signature, i.e. strong Balmer absorption lines. As is shown in Fig.~\ref{postspec}, stellar post-starburst spectra, mixed with emission line spectra,  are seen in 3 cases.  The post-starburst signatures start to develop a few 10 Myr after the onset of a starburst and, if the starburst decays, achieve a maximum after a few 100 Myr and are gone after not much more than 1 Gyr. The conditions for leakage would be improved in such cases if the starburst that happened some time ago was strong enough to expel and heat the gas in the main body and open channels towards the IGM through which the ionizing photons could escape. With velocities of the order of a few 100 km$s^{-1}$ \citep{1990ApJS...74..833H} the winds need $\sim$10 Myr to clear the central few kpc:s. During this epoch the 'post-starburst' population is just about to emerge while hot stars are still present and can leak out ionizing radiation. We expect to detect the post-starburst population at a younger age than normal if the emission lines are weakened due to leakage.


In Fig.~\ref{wha} we display \xwha ~ maps. What is characteristic of all but one galaxy (no. 8) is the chaotic structures in \xwha, strongly deviating from the distribution of the red stellar continuum and in most cases also the \ha emission. It essentially reflects the irregular distribution of the stellar population of different ages. Apparently, most of the galaxies are not in equilibrium. The deviation from equilibrium favours leakage since young star clusters may be able to separate from the gas and during brief period be located in regions where the column density of the absorbing neutral gas in some directions may be significantly lower than normal. This is likely to occur if the dynamical instabilities cause the clusters to migrate outwards on short timescales or if the galaxy is exerted to ram pressure stripping.

\subsection{Ram pressure stripping}

A few of our targets are residing in small groups of galaxies in obvious interaction with the neighbours, and one may consider the importance of ram pressure stripping as the galaxy moves through the common hot intergalactic medium.  Models of stripping indicate that star formation can be increased significantly during the process \citep{2008A&A...481..337K,2009MNRAS.399.2004M}. On a fairly short time scale, $\sim$ 100 Myr, the gas on the side facing the wind may be severely depleted \citep{2008A&A...481..337K} while strongly ionizing stars are still alive in the low density wake immediately behind. It thus seems possible that both the gas removal and the star formation triggering by the motion through the intergalactic medium could favour leakage. Does it have any significant effect in groups? Certainly the \ha morphology of some of our targets remind of the cometary or tadpole \citep{1996AJ....112..359V} structure of wakes predicted from models of galaxies being stripped \citep[see e.g. Fig 11. in ][]{2009A&A...499...87K}.

In a recent paper,  \citet{2011ApJ...738..145F} find that groups of galaxies of the size about a magnitude more massive than the local group can contain intergalactic clouds with densities 10$^{-3}<n< $10$^{-4}$ and velocity dispersions of a few 100 kms$^{-1}$. In these cases, the ram pressure is sufficiently strong to completely remove the gas from galaxies in the mass range $\sim 10^{6-7}{\cal M}_\odot$ but is probably insufficient for masses above 10$^9 \cal M_\odot$. As discussed below, our galaxies are in the intermediate range and may be severely affected. Two cases where ram pressure effects seem be active, as seen from their morphology,  were discussed above and displayed in Fig. \ref{br_wha}. In an interesting comparative study of Virgo vs. field galaxies,  \citet{2006AJ....131..716K} find that the ratio between the \ha scale lengths and the R-band scale lengths of spiral galaxies were systematically different between the two samples. Field galaxies had a mean ratio of $h_{H\alpha}/h_R$=1.14$\pm$0.07, while the Virgo galaxies had an \ha scale length which was 20\% lower than the R-band scale length. They interpret these results as an environmental effect. In the Virgo cluster, ram pressure stripping is efficiently removing the gas from the galaxies as they move through the cluster. Some of the Virgo galaxies show a strong displacement of ionized gas in a direction opposite to the centre direction. These also show an increased star formation activity on the side facing the centre. A similar process is going on in the Large Magellanic Cloud.  

Can we apply the results obtained by Koopmann et al. on our sample? The sample differs in mean luminosity but about half of their galaxies have absolute magnitudes in the range -18.5$>$M$_B>$-19.5, which encompasses about 50\% of our sample. Koopmann et al.  also concluded that they found no trends in $h_{H\alpha}/h_R$ with neither luminosity nor Hubble type. Most of our galaxies show disk structure but due to the fact that they tend to appear disturbed we refrain from making a proper bulge-disk separation and we assume that we view the systems face on. We choose the brightest part in the \ha continuum as the centre and then derive the luminosity profiles in circular strips. Then we inspect the profiles and fit a linear relation in $R$-$\mu$ to the nearly exponential part. In all cases we leave out the central region, corresponding to a radius of 3-4\arcsec.  In Table 2 we list the scale lengths in \ha and the \ha continuum, excluding no. 2, which is of an M51 companion type. The continuum wavelength ($\sim$ 6014\AA) is fairly close to the R band. The R band also contains \ha but in our case the contribution is less than 10\% to the R-band so we can more or less directly compare our scale lengths to the results by Koopmann et al.. In all cases the \ha scale length is shorter than the continuum scale length. In the mean, $<h_{H\alpha}/h_{H\alpha -cont}>$ = 0.71, i.e. the \ha scale length is 30\% smaller. Thus the effect is stronger than in the Virgo galaxies and signals that ram pressure stripping and possibly star formation triggering may be active. We find this to be an additional support for leakage in our galaxies.


\section{Discussion}
\subsection{Local versus global leakage}
In current simulations of Lyman continuum leakage from galaxies \citep[e.g.][]{2008ApJ...672..765G,2010ApJ...710.1239R}, the escape fraction is typically defined at the virial radius $r_\mathrm{vir}$ of the dark matter halo. Since $r_\mathrm{vir}$ is much larger than the visible extent of any galaxy, this leads to a slight problem when attempting to indirectly assess the escape fraction from observational data such as ours. There is simply no way that we can trace the light profiles all the way out to $r_\mathrm{vir}$, and if the broadband colours and H$\alpha$ equivalent widths were very different once integrated out to this radius, this would indicate that the leakage is local rather than global. In other words, Lyman continuum photons may be escaping from the central regions of our targets, yet be absorbed before reaching the intergalactic medium.

To estimate the virial radii of our objects, we assume a stellar mass-to-light ratio in the range ${\cal M}_\mathrm{stars}/L_g\sim 0.01$--0.1 (in solar units) for the sample (as derived by mixing burst and post-burst populations as in Sect.~\ref{section:model}), and a gas mass fraction of $f_\mathrm{gas}\approx 0.5$. After allowing for a small amount of extinction ($A(g)\leq 1.0$ mag), this implies a mass in detected baryons of $\sim 10^{8-9}{\cal M}_\odot$ for our targets. Using the baryonic Tully-Fisher relation \citep{2010ApJ...708L..14M}, we can then estimate the total dark matter masses of our objects. We arrive at ${\cal M}_{500}\sim 10^{10-11}{\cal M}_\odot$, where ${\cal M}_{500}$ represents the mass inside which the average density is 500 times the critical density of the Universe. This can be converted into a virial mass and a virial radius using relations in \citet{1998ApJ...495...80B} and \citet{2001MNRAS.327.1353K}. The result is that the virial radii of our target galaxies are likely to be $r_\mathrm{vir}\approx$ 50--100 kpc at $z\approx 0.03$. In terms of angular sizes, the virial diameters are likely to subtend angles of $\approx 3$--$6\arcmin$ in the sky at $z\approx 0.03$.

The SDSS apertures ($3\arcsec$ across) used when selecting our Lyman continuum leakage candidates are admittedly much smaller than this,
and although our $B$-band and H$\alpha$ images can trace the luminosity profiles somewhat further out, the signal is inevitably lost in the sky noise at $\mu_B \sim$ 28.6 (3$\sigma$ arcsec$^{-2}$), corresponding to typical radii of not more than a few tenth of an arcminute. However, the shape of our surface brightness profiles ($h_{H\alpha} \sim$ 1kpc) suggest that the flux from the faint outskirts that we miss contributes no more than ~10\% to the total. Hence, the LyC escape fractions inferred by our method likely represent the global values, and should therefore be directly comparable with the results from simulations if there is negligible absorption in the halo.

\subsection{Evolution with redshift}
\label{section:redshift}
As we pointed out in Sect. 3.3, only $\sim$ 0.1\% of the galaxies at the redshift we are exploring fulfil our selection criteria. If these criteria constitute a necessary condition for a significant leakage it means that leakage from such galaxies is completely unimportant for the ionization balance in the local universe. If the same conditions hold at high redshifts the leaking Lyman photons would contribute insignificantly to the reionization. How do we expect the conditions to change with redshift? As the morphology of our targets tell us, there is a tendency for disturbed morphologies with indications of mergers or interactions with neighbouring galaxies or the intergalactic (intragroup or intracluster) medium. Some show  tadpole structures, morphologies interpreted as a result of mergers or ram pressure sweeping.  How frequent are these objects? Locally, $\sim$0.2\% of the Kiso UV excess galaxies \citep{1993PNAOJ...3..169T} show tadpole structures \citep{2012arXiv1203.2486E}, i.e. the same amount as the galaxies that obey our search criterion. The number of tadpole galaxies increases dramatically with redshift and in the UDF the frequency among bright galaxies is $\sim$ 10\% \citep{2012arXiv1203.2486E}. The same authors find that the increase in SFR in the tadpoles is even more dramatic, of the order of two magnitudes from z=0 to 3, for the mass range we are dealing with. This is much higher than the increase in relative gas mass fraction. An increase in gas mass fraction clearly makes it less favourable for LyC escape but this may be compensated for by the increase in SFR in the sense that the fraction of ionized gas and the importance of SN superwinds increase.  If the lifetime of the event when the leakage can take place is short (e.g. $\leq$ a few 10 Myr during a stripping process), then the detection rate of the cases where the leakage is high increases with the volume density of the objects leaking. So the increased escape fraction can partly be explained by the observed increase of galaxies exposed to ram-pressure stripping and mergers and the increased SN activity. From these considerations we would expect a gradual change in LyC escape with redshift. In Fig.~\ref{fesc} we see a dramatic increase in the LyC escape fraction with redshift but at $z \sim$1 we only have upper limits. Why is it so? One possible explanation, related to technical aspects, was discussed in Sect.~\ref{section:escape}. We also noted the possibile importance of dust attenuation. Although the dust attenuation of the galaxies at redshift of 0.03 is not 'serious' in the optical region, the LyC may be attenuated with up to a factor of 10. Studies of the evolution of the dust content in the universe gives interesting results. \citet{2011ApJ...730....8H} and \citet{2012MNRAS.422..310G} have investigated the correlation between \lya escape and dust content with redshift and find a tight relation showing that \lya escape increases as the dust content in galaxies decreases with redshift. This is of particular interest since \lya and LyC escape should be related if the leakage is from a density bounded region. \citet{2012A&A...539A..31C} find that the dust attenuation increases from zero redshift, reaching a peak at $z \sim$1 and then decays to a level similar to today at $z \sim$3. All these effects combined can help to explain the observations summarized in Fig.~\ref{fesc}. The importance of sub-L$^*$ galaxies in the reionization process is further strengthened by the fact the dust content is a steep function of stellar mass, preferentially allowing more generously for leakage in low mass galaxies \citep{2012MNRAS.421.2187S}.

\section{Summary and Conclusions}

In this paper we suggest that the lack of detection of leaking local galaxies could, at least partly, be due to a selection bias. Leaking galaxies, by definition, should use less photons for ionization than normal star forming galaxies and therefore the emission lines should be weaker. However, all searches so far have been focused on galaxies with strong emission. Here we select galaxies which have fairly weak Balmer emission lines but extremely blue colours, typical of a galaxy dominated by young stars but where the ionizing photons are leaking out. Using our spectral evolutionary model we select galaxies from the SDSS with the properties of a leaker. We select eight galaxies at a redshift of $z$ = 0.03 for imaging in broadband B and \xha. Although the sample is small we find some characteristic properties that deviate from those of normal star forming galaxies and support LyC leakage. These are

\begin{itemize}

\item Among the eight target galaxies, six show clearly disturbed morphologies, signs of interactions, mergers or ram-pressure stripping. Close neighbours that also show morphological distorsions are found in the environments of four-five galaxies.

\item In most of the galaxies we find an off-center located population of young stars that, in a colour map, seems to connect to the halo. This shows that the conditions for leakage are improved since the hot stars are located where the gas column density is low. In all cases the \ha scale length is significantly (30\%) smaller than the scale length of the stellar continuum. This is an indication of ram pressure stripping of the gas and explains our estimated low density in the ionized gas and some morphological features.

\item Young post-starburst signatures are seen in the stellar spectra of three of the eight galaxies. This may partly be an artefact caused by a strong leakage that makes the spectrum look as if the starburst has ceased. It indicates however that a major star formation epoch started more than $\sim$ 10 Myr ago. During this epoch winds from massive stars have had time to clear the sight in the central region.

\end{itemize}

In summary, we find that a plausible explanation for why there are so few detections of  Lyman continuum leaking galaxies in the local universe is that we have been ignorant of their presence in our search. If we instead of looking for leakage from strong starburst galaxies more focus on galaxies with blue continuae and weak emission lines, our investigation of this small template sample shows that we may be on the right track. In a truly leaking galaxy, the radiation manages to reach outside the virial radius. Although we cannot map the whole volume out to this radius, our estimate is that the amount of radiation absorbed outside what we can observe is insignificant and that the galaxies most likely are leaking. There seems to be specific conditions at hand that favour leakage. These are the involvement in mergers and interactions with neighbours, interaction with the intergalactic medium via ram pressure stripping and a recent major starburst as a precursor to the present young leaking stellar population. 

An additional factor that seems to play a major r\^ole over cosmic time scales is the dust content. \lya emission has been found to anti-correlate with an increasing dust content in galaxies over time. The redshift dependence of LyC leakage indicates that dust may also contribute significantly to the possibilities for the LyC photons to escape.

\begin{acknowledgements}
We wish to thank the anonymous referee for a wealth of constructive comments and corrections that resulted in a significantly improved quality of the paper.
Bergvall and Zackrisson acknowledge support from the Swedish National Space Board and the Swedish Research Council.
\end{acknowledgements}

\bibliographystyle{aa} 
\bibliography{bergvall_1303} 

\appendix
\section{Spectral modelling}

The models are based on spectra from the  \citet{2001A&A...375..814Z} spectral evolutionary models library. The stellar populations have a Salpeter initial mass function and a mass range of 0.08-120 \sma. The metallicities range from 5\% solar to solar. A nebular component  based on CLOUDY \citep{1996hbic.book.....F, 1998PASP..110..761F} is also included, making it possible to model galaxies with different amounts of leakage. In the basic setup the models comprise a set of  spectra of stellar populations of different ages and {\sl modes} of star formation histories. The {\sl mode} can be either constant star formation over a certain time period or an exponentially declining star formation $SFR \propto e^{-t/\tau}$, where t is the age and  $\tau$ is the decay timescale. 

In the modelling of the SDSS spectra in the present investigation we assume that the stellar population consists of a mixture of a young and an old population as if a burst of star formation recently took place in an old passive galaxy. To mimic the conditions at reionization, the young population cannot be too old so we assume a burst duration of 10 Myr and then an abrupt shut down of star formation. The metallicities are 40\% solar. We mix these components in different proportions and only consider the case of constant star formation for the young component.  The spectrum can be described as as a sum of young and old stars plus a nebular component, where the young stellar component and the nebular components are age dependent: $F_{\lambda}(t)=F_{\lambda, young}(t)+F_{\lambda, old}+F_{\lambda, H II}(t)$. The old component is assumed to have experienced a short initial burst of star formation 10 Gyr ago, followed by passive evolution. We chose this age to agree with the cosmic look back time at peak cosmic  SFR \citep{2006ApJ...651..142H}. Observations of local dwarf galaxies also support the formation of most of the stars at z$>$1 \citep{2011ApJ...739....5W}. The star formation activity at a specific age may conveniently be expressed by the $b-parameter$, the ratio of the present SFR and the mean past SFR, preceding the start of the burst: $b$ = SFR/$<$SFR$>$.

The young and old components are mixed in different proportions and the relative mass fraction of young and old stars as function of age of the young component can be expressed as

\begin{equation}
\frac{{\cal M}_{young}}{{\cal M}_{old}} = b\frac{t_{young}}{t_{old}}
\end{equation}

where $t_{young}$ and $t_{old}$ are the ages of the young and old population respectively. We calculate a grid of models where we vary two parameters - the $b$-parameter and the escape fraction,  $f_{esc}$. In the code, escape fractions above $f_{esc}=0$ can be simulated by numerically reducing the contribution from  the nebular component with a suitable amount. Emission lines and the nebular continuum will be affected the same way. If $F_{H\alpha_0}$ refers to the \ha flux when there is no leakage of ionizing radiation and $F_{H\alpha}$ is what we measure when there is leakage we find

\begin{equation}
f_{esc}=(1-F_{H\alpha}/F_{H\alpha_0})=(1-F_{\lambda,nebC}/F_{\lambda,nebC_0})
\end{equation}

where $F_{\lambda,nebC}$ and $F_{\lambda,nebC_0}$ is the flux density of the nebular free-free and free-bound continuum emission with and without leakage.
Consequently we may derive the equivalent width of \ha from

\begin{equation}
EW(H\alpha) = \frac{F_{H\alpha_0}(1-f_{esc})}{F_{\lambda,old}+F_{\lambda,young}+F_{\lambda,nebC_0}(1-f_{esc})}
\end{equation}

where $F_{\lambda,old}$ and $F_{\lambda,young}$ are the flux densities of the continuum under  under \ha with contributions from the old stars and the young stars respectively. When we follow the evolution of the spectrum with time, the galaxy will experience a period where 20\AA $ <$\wha$<$ 200\AA. The larger the escape fraction, and assuming the same $b$-parameter, the lower will be the age of the young population during the epoch that this criterion will be fulfilled.  When one considers the characteristics of spectra of leaking galaxies one should remember that, since $F_{\lambda,nebC}$ dominates the continuum under \ha in young starbursts and \ha scales linearily with $F_{\lambda,nebC}$, \wha will initially drop only slowly with increasing $f_{esc}$. 

\end{document}